\documentclass[aps,prx,twocolumn,amsmath,amssymb,superscriptaddress,notitlepage,longbibliography]{revtex4-1}

\usepackage[colorlinks=true,citecolor=blue,linkcolor=blue,hypertexnames=false]{hyperref}
\usepackage{amsmath}
\usepackage{dsfont}
\usepackage{hyperref}
\usepackage{textcomp}
\usepackage{tabularx}
\usepackage[compat=1.1.0]{tikz-feynman}
\usepackage{esvect}

\newcommand{\gdbar}[1]{\overline{\overline{g}}}

\def \k{{\mathbf k}}
\def \br{{\mathbf r}}
\def \bk{{\mathbf k}}
\def \bq{{\mathbf q}}
\def \q{{\mathbf q}}

\def \r{{\mathbf r}}

\def \G{{\mathbf G}}

\def \r{{\mathbf r}}

\def \be{\begin{equation}}
\def \ee{\end{equation}}

\begin{document}

\title{Spin-triplet superconductivity from inter-valley Goldstone modes in magic-angle graphene}

\author{Vladyslav Kozii}
\affiliation{Department of Physics, University of California, Berkeley, CA 94720, USA}
\affiliation{Materials Sciences Division, Lawrence Berkeley National Laboratory, Berkeley, CA 94720, USA}
\affiliation{Department of Physics, Carnegie Mellon University, Pittsburgh, Pennsylvania 15213, USA}

\author{Michael P. Zaletel}
\affiliation{Department of Physics, University of California, Berkeley, CA 94720, USA}
\affiliation{Materials Sciences Division, Lawrence Berkeley National Laboratory, Berkeley, CA 94720, USA}

\author{Nick Bultinck}
\affiliation{Department of Physics, University of California, Berkeley, CA 94720, USA}
\affiliation{Department of Physics, Ghent University, 9000 Ghent, Belgium}

\date{\today}
\begin{abstract}
We consider magic-angle graphene in the doping regime around charge neutrality and study the connection between a recently proposed inter-valley coherent insulator at zero doping and the neighboring superconducting domes. The magic-angle graphene continuum model has an emergent U$(1)$ valley-charge conservation symmetry, and an emergent SU$(2)$ symmetry corresponding to opposite spin rotations in the two valleys. The inter-valley coherent insulator spontaneously breaks both these emergent symmetries, and as a result has four Goldstone modes which couple to doped charge carriers. We derive the effective interaction mediated by the Goldstone modes, and study its role in electron pair formation. The SU$(2)$ Goldstone modes generate a ferromagnetic interaction, which is attractive in spin-triplet pairing channels and repulsive in spin-singlet channels. From a weak-coupling BCS calculation, we find the leading superconducting instability in the $p-$wave channel.
\end{abstract}
\maketitle

\section{Introduction}

The experimental discovery of superconductivity in Magic-Angle Twisted Bilayer Graphene (MATBG)~\cite{PabloSC} has spurred a tremendous interest in developing a theoretical understanding of the underlying pairing mechanism~\cite{Guinea2018,WuMacDonaldMartin,LianBernevig,YiZhuangVishwanath,Peltonen,Choi,XuBalents,Bitan19,IsobeYuan,LinNandkishore1,LinNandkishore2,WuHwangDasSarma,Sharma,KoziiIsobe,ClassenChubukov,Dodaro,Venderbos,RayJung,Rademaker,Kennes,Liu,Guo,Spalek,Stauber,FengCheng,Sherkunov,SkyrmionSC}. At present, the microscopic origin of superconductivity in MATBG is still under debate. For example, it is not clear whether electron pairing is the result of phonon exchange, or whether it is driven by a more exotic mechanism coming from Coulomb repulsion. It is also equally unclear whether the superconducting domes are in any way related to the correlated insulating phases which are observed in transport experiments at certain integer fillings~\cite{PabloMott,Dean-Young,efetov,Sharpe,YoungAH} (signatures of these insulating phases are also seen in spectroscopic measurements~\cite{CaltechSTM,ColumbiaSTM,PrincetonSTM,RutgersSTM,Wong,Shahal}). This last question was addressed in more detail in two recent experimental works~\cite{EfetovScreening,YoungScreening}, where superconducting domes in the doping regimes around two electrons or holes per moir\'e unit cell ($\nu=\pm 2$) were observed without any signature of a correlated insulator. 

In Refs.~\cite{efetov,EfetovScreening} superconducting domes were also observed next to the charge neutrality point. In Ref.~\cite{EfetovScreening} it was found that these domes appear only when the distance between the MATBG device and the gates is large enough, i.e., when screening by the metallic gates is sufficiently weak. This observation suggests that Coulomb repulsion plays an important role for the origin of superconductivity near charge neutrality. Interestingly, in the same devices insulating behavior is also observed at charge neutrality~\cite{efetov,EfetovScreening} (a charge gap was also observed in the tunneling experiments of Ref. \cite{Wong}).

In Ref.~\cite{KIVC} it was proposed that the insulating behavior of magic-angle graphene at charge neutrality is the result of an inter-valley coherent order which develops at zero temperature. This order implies that the electron system spontaneously breaks several continuous symmetries, and therefore hosts Goldstone modes. The inter-valley coherent insulator was dubbed the K-IVC (Kramers inter-valley coherent) insulator \cite{KIVC}, because it is invariant under an emergent spinless Kramers time-reversal symmetry.

In this work, we take the K-IVC insulator of Ref.~\cite{KIVC} as the starting point for a study of superconductivity in MATBG near charge neutrality. In particular, we investigate the potential role of the Goldstone modes in the formation of Cooper pairs. More concretely, we study how the attractive interaction mediated by the exchange of the Goldstone modes, taken together with the screened Coulomb interaction, can give rise to the superconducting instabilities of the doped insulator near charge neutrality. Importantly, we find that the density of states of the doped K-IVC insulator is significantly smaller than  that of the ``bare'' (non-interacting) nearly flat bands of MATBG at the magic-angle ~\cite{Morell2010,Santos,BM}, which allows us to treat the problem within a weak-coupling approach. We note that the role of inter-valley Goldstone modes for superconductivity has also been discussed previously in Ref. \cite{PoZou}, where the authors considered a phenomenological inter-valley coherent insulator in MATBG.

Our main result is that the exchange of inter-valley Goldstone modes generates an attractive interaction in spin-triplet Cooper channels only. This can be understood as follows. At low energies, magic-angle graphene has an emergent valley U$(1)$ symmetry, and to a very good approximation also an additional emergent SU$(2)$ symmetry which physically corresponds to opposite spin rotations in the two valleys. Inter-valley coherent states spontaneously break both these emergent symmetries, and as a result have four linearly dispersing Goldstone modes.  Three of these Goldstone modes -- those corresponding to the broken generators of SU$(2)$  -- generate an effective ferromagnetic interaction between doped electrons. This interaction is attractive in the $p$-wave and repulsive in the $s$-wave pairing channels. The Goldstone mode generated by U$(1)$ symmetry breaking is attractive in both $s$-wave and $p$-wave pairing channels. We find that the overall interaction mediated by the exchange of all four Goldstone modes pairs electrons in a spin-triplet $p$-wave channel and is repulsive in spin-singlet channels. 


We perform a weak-coupling BCS-type calculation and find a superconducting instability in the $p-$wave channel. The corresponding dimensionless pairing strength $\lambda$ grows as a function of the screening length $D$, which is consistent with the observations of Ref.~\cite{EfetovScreening}. However, the value $\lambda = 0.12$ obtained at the experimentally relevant screening length $D = 15$ nm value is too small to explain the observed superconducting transition temperature $T_c \approx 0.3$ K~\cite{EfetovScreening}. We expect that our naive BCS calculation underestimates $T_c$. More sophisticated calculations which allow for a frequency-dependent pairing gap can potentially produce higher critical temperatures. Our results do not incorporate the retardation effects that are crucial to weaken strong Coulomb repulsion in conventional metals. The absence of retardation poses a serious problem in the study of superconductivity in low-density materials with a small Fermi energy~\cite{RuhmanLee,Kozii,RuhmanFernandes}, and we therefore expect that taking retardation into account will lead to significantly higher values of $T_c$.

As found in Ref.~\cite{KIVC}, the inter-valley coherent order at charge neutrality gets destroyed when the MATBG is aligned with the hexagonal Boron-Nitride substrate. The pairing instabilities discussed in this work (which rely on Goldstone modes resulting from inter-valley coherent order) would then disappear together with it. This could explain why no superconductivity is seen in the substrate-aligned devices of Refs.~\cite{Sharpe,YoungAH}. 

The remainder of the paper is organized as follows. In Sec.~\ref{sec:KIVCreview}, we start by reviewing the essential properties of the KIVC insulator found in Ref.~\cite{KIVC}. The properties of the Goldstone modes of the KIVC insulator are discussed in Sec.~\ref{sec:GSmodes}. In Sec.~\ref{sec:dopedKIVC}, we derive the coupling between electrons in the conduction bands of KIVC state and these Goldstone modes (for concreteness, we focus on electron doping). The effective interaction between the electrons mediated by the exchange of Goldstone modes is derived in Sec.~\ref{sec:SC}. In Sec. \ref{sec:instabilities}, the superconducting instabilities in the presence of both the Coulomb interaction and the Goldstone mode-mediated interaction are examined. We end with a discussion and outlook in Sec.~\ref{sec:discussion}. In the Appendices, we provide additional details on how to calculate the Goldstone mode propagator at charge neutrality, and on how to derive an effective low-energy theory for the doped electron system and the Goldstone modes. We also give a more thorough discussion of the analysis of the superconducting instabilities. 

\section{The K-IVC insulator at charge neutrality}\label{sec:KIVCreview}

Our starting point is the Kramers intervalley coherent (K-IVC) insulator, which was found to describe the ground state of magic-angle graphene at charge neutrality in Ref. \cite{KIVC} (similar states were also discussed previously in Refs.~\cite{PoZou,KangVafekPRL}). Evidence for intervalley coherence in MATBG at charge neutrality was also found in quantum Monte Carlo \cite{DaLiaoFernandes}. We write the mean-field Hamiltonian of the K-IVC insulator as follows:

\begin{equation}\label{ham}
    H = H_0 + H_{\Delta} = \sum_{\k}c^\dagger_\k\left[h_0(\k) + \Delta(\k)\right]c_\k\, ,
\end{equation}
where the electron operators $c^\dagger_\k$ are defined in the band basis of the Bistritzer-MacDonald (BM) model of MATBG \cite{BM}. We consider a model which is obtained by projection into the 8 narrow bands of MATBG, such that each electron operator carries a BM label $a$, together with valley and spin quantum numbers $\tau\in\{+,-\}$ and $s \in\{\uparrow,\downarrow\}$. For our numerical simulations, we keep six bands per spin and valley, corresponding to the two narrow bands and the first two remote bands above and below charge neutrality. As these are only bands that lie in the energy window $[-100 \text{ meV},100 \text{ meV}]$ of the BM Hamiltonian, and because the Coulomb interaction energy at the moir\'e scale is $\sim 60$ meV, we expect that other BM bands at higher energies will not be important for the strong interaction physics in the flat bands. This assumption was verified at the mean-field level in Ref.~\cite{KIVC}.

\begin{figure}
\includegraphics[scale=0.32]{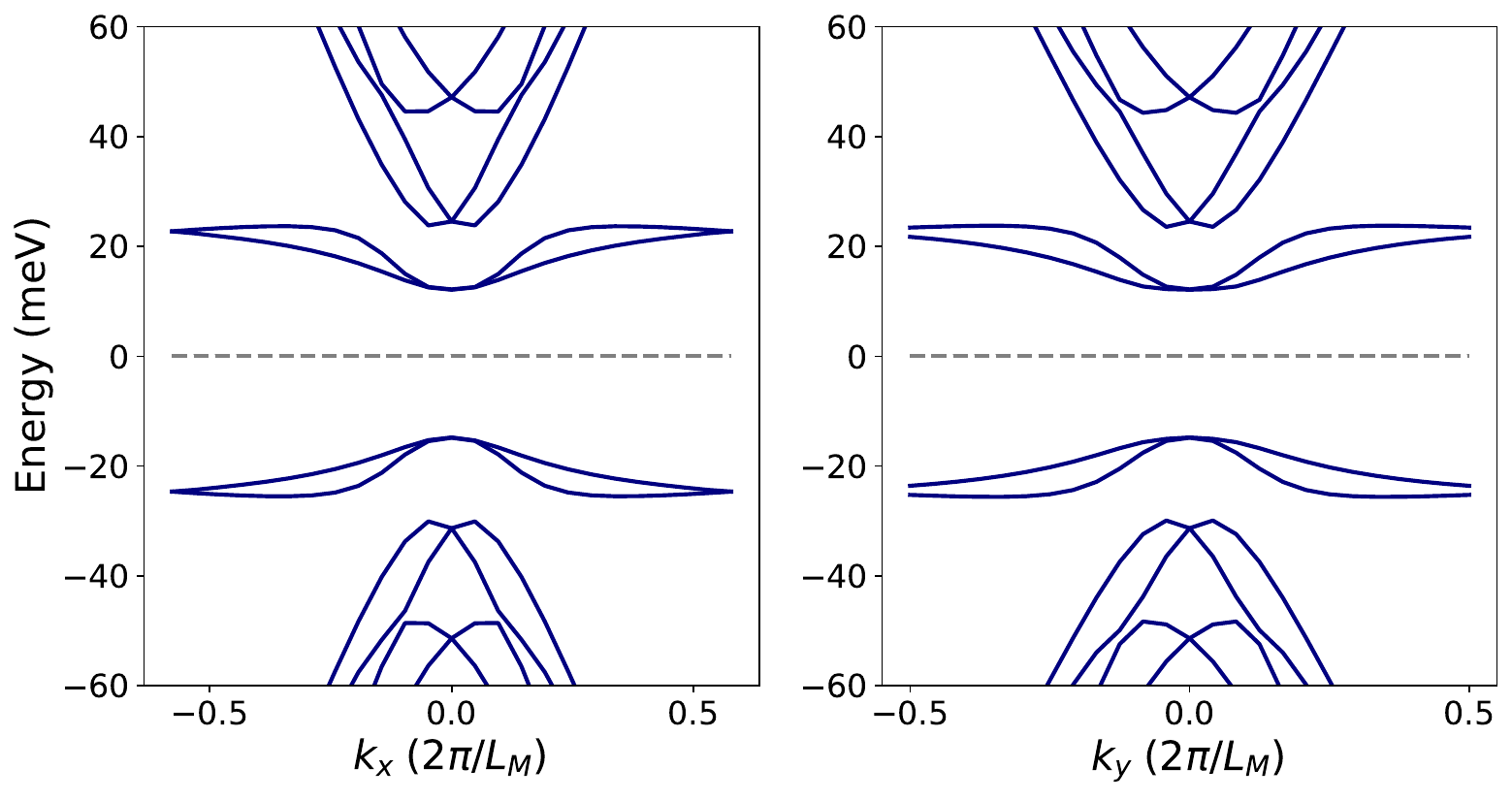}
\caption{Self-consistent mean-field band structure of the K-IVC state at charge neutrality along two cuts through the mini-BZ, one in the $x-$direction (left panel) and one in the $y-$direction (right panel). The parameters which were used in the BM Hamiltonian are $\theta = 1.09^\circ$, $w_1=110$ meV and $w_0/w_1 = 0.75$, where $w_0$ ($w_1$) is the sublattice diagonal (off-diagonal) interlayer tunneling strength. For the interaction, a dual-gate screened Coulomb interaction with gate distance $D=15$ nm and dielectric constant $\epsilon = 10$ was used. For more details, see Ref.~\cite{KIVC}.}\label{fig:cuts}
\end{figure}

The term $H_0$ in Eq. \eqref{ham} represents the symmetric part of the Hamiltonian, and $H_{\Delta}$ is the order parameter contribution. The order parameter introduces coherence between electrons from different valleys, and is therefore off-diagonal in valley-space, i.e. $\tau^z\Delta(\k)\tau^z = - \Delta(\k)$, where $\tau^z$ is the third Pauli matrix acting on the valley indices. As a result, the order parameter signals the spontaneous breaking of the valley U$(1)$ symmetry (denoted as U$_V(1)$) of the interacting BM model, which acts as

\begin{equation}
    \text{U}_V(1) : \;\;c^\dagger_{\tau,s,a,\k} \rightarrow e^{i\tau\phi}c^\dagger_{\tau,s,a,\k}.
\end{equation}
In Fig.~\ref{fig:cuts}, the mean-field band spectrum of the K-IVC insulator obtained from self-consistent Hartree-Fock is shown along two orthogonal cuts through the mini-Brillouin zone (mini-BZ). Note that on top of the two-fold spin degeneracy, the K-IVC band spectrum has additional degeneracies at time-reversal invariant momenta. This is the result of a \emph{spinless} Kramers time-reversal symmetry $\mathcal{T}'$, which acts as

\begin{equation}
\mathcal{T}': \;\;c^\dagger_{\tau,s,a,\k}  \rightarrow e^{i\tau \pi/2}c^\dagger_{-\tau,s,a,-\k}\, ,\;\;\; i\rightarrow -i, \label{Tsymm}
\end{equation}
and satisfies $\mathcal{T}'^2 = -\mathds{1}$.
The Kramers time-reversal symmetry $\mathcal{T}'=\tau^y\mathcal{K}$ (where $\mathcal{K}$ denotes complex conjugation) is a combination of the conventional spinless time-reversal $\mathcal{T} = \tau^x\mathcal{K}$, and a $\pi/2$ valley U(1) rotation.

Finally, let us comment on the spin properties of the K-IVC state. Without loss of generality, we can take the K-IVC insulator to be spin-singlet \cite{KIVC}, i.e. $\left[\Delta(\k)\right]_{(\tau,a,s),(\tau',b,s')} = \left[\Delta'(\k)\right]_{(\tau,a),(\tau',b)}\delta_{s,s'}$. An important point is that the leading part of the interacting BM model has as an enhanced SU$_+(2)\times$SU$_-(2)$ spin rotation symmetry, where SU$_{\tau}(2)$ corresponds to spin rotations in valley $\tau$. This enhanced symmetry is broken down to the conventional SU$(2)$ spin rotation symmetry by an intervalley exchange coupling, which is of the order $\sim 0.1$ meV. As is common in studies of MATBG, we ignore this small intervalley exchange coupling. Because the K-IVC order introduces coherence between electrons from different valleys, it spontaneously breaks the subgroup of SU$_+(2)\times$SU$_-(2)$ corresponding to opposite spin rotations in the two valleys.  In this work, we consider without loss of generality the spin-singlet K-IVC state, which has an unbroken global spin rotation symmetry (corresponding to the \emph{same} spin rotation in the two valleys).

\section{Goldstone modes of the K-IVC insulator}\label{sec:GSmodes}

The K-IVC insulator spontaneously breaks several continuous symmetries, such as the valley U$(1)$ symmetry, and an SU$(2)$ symmetry corresponding to opposite spin rotations in different valleys. The corresponding broken generators are given by

\begin{equation}
    \tau^z,\;\tau^zs^x, \;\tau^zs^y,\; \tau^zs^z\, ,
\end{equation}
where $s^i$ are the Pauli matrices acting on spin indices. In the absence of Lorentz symmetry, the number of Goldstone modes associated with the spontaneously broken symmetries can be smaller than the number of broken generators. In Refs. \cite{Hidaka,Watanabe1,Watanabe2}, it was shown that in general, the number of Goldstone modes is given by 

\begin{equation}
    n_{GS} = n_{BS} - \frac{1}{2}\text{rank}\rho\, ,
\end{equation}
where $n_{BS}$ is the number of broken symmetry generators. The real, anti-symmetric matrix $\rho$ is defined using the local broken charges $q_\mu$ (for translationally invariant systems) as

\begin{equation}
    \rho_{\mu\nu} =   -i\langle\left[q_\mu,q_\nu\right]\rangle\,,
\end{equation}
where the expectation values are taken with respect to the ground state wavefunction. For the K-IVC insulator, we can take

\begin{equation}
    q_\mu = \frac{1}{2N}\sum_\q\sum_\k c^\dagger_{\k+\q} \tau^zs^\mu c_\k\, ,
\end{equation}
with $\mu = 0,x,y,z$ and $s^0 = \mathds{1}$. Using these expressions, we find that $[q_0,q_\mu] =0$, and

\begin{equation}
    [q_i,q_j]=i\epsilon_{ijk}\frac{1}{2N}\sum_\q\sum_{\k} c^\dagger_{\k+\q}s^kc_\k\,,
\end{equation}
with $i,j,k\in \{x,y,z\}$. The spin rotation symmetry of the K-IVC state (recall that we consider a spin-singlet K-IVC state, which is invariant under applying the same spin rotation in both valleys) implies that $\langle [q_i,q_j]\rangle = 0$. We thus arrive at the conclusion that $\rho = 0$, which implies that the K-IVC insulator has 4 Goldstone modes.

\begin{figure}
    \centering
    \includegraphics[scale=0.42]{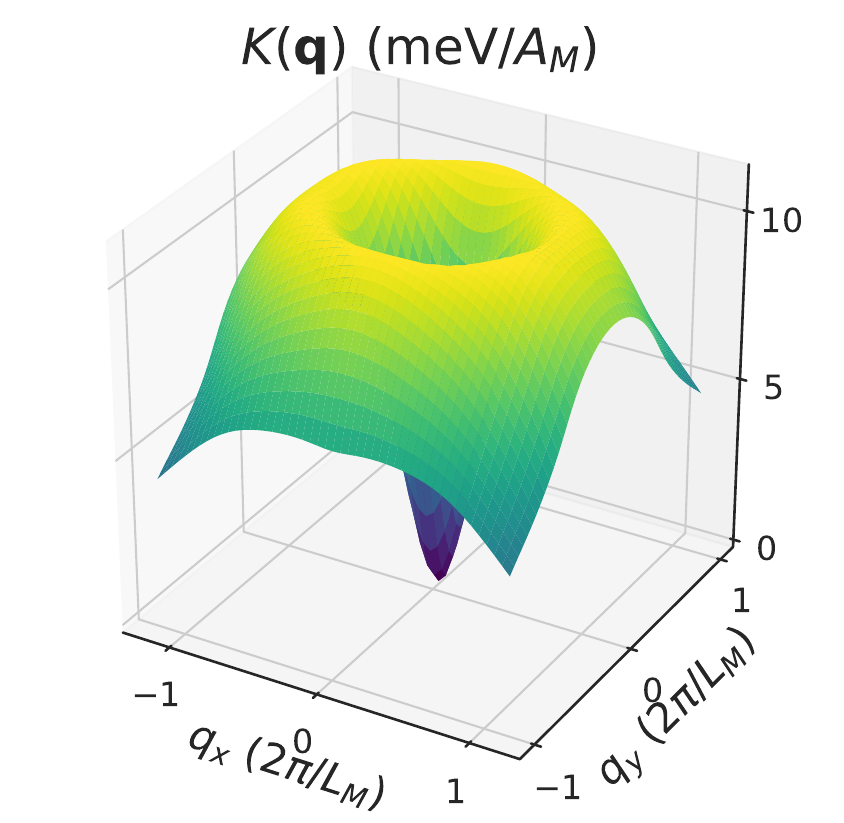}
    \caption{$K(\q)$ at charge neutrality, see Eq.~\eqref{GSprop}.}
    \label{fig:Kq}
\end{figure}

The Goldstone modes can be described using the following effective Euclidean action: 

\begin{equation}\label{GSaction}
S_G = -\frac{1}{2} \int\frac{d\omega \text{d}\q}{(2\pi)^3} \phi_\mu(i\omega,\q)D^{-1}(i\omega,\q)\phi_\mu(-i\omega,-\q)\, ,
\end{equation}
where $\phi_\mu(i\omega,\q)$ are the Goldstone fields, summation over the repeated index $\mu$ is implicit, and

\begin{equation}\label{GSprop}
    D^{-1}(i\omega,\q) = \chi_s(i\omega)^2 - K(\q)\, .
\end{equation}
In Appendix~\ref{app:stiffness} we calculate $K(\q)$ from the mean-field K-IVC band structure. The result of this calculation is plotted in Fig. \ref{fig:Kq} (for $\q \lesssim 2\pi/L_M$, where $L_M\approx 13$~nm is the moir\'e lattice constant). From the long wavelength part of $K(\q)$, we obtain the stiffness of the K-IVC state $\rho_s$ as $K(\q) = \rho_s\q^2+\mathcal{O}(\q^4)$. Numerically, we find that $\rho_s \approx 4$ meV, which agrees with the value obtained in Ref.~\cite{SkyrmionSC} via a different method. The `compressibility' $\chi_s$ of the Goldstone modes is not important for our analysis below. The low-energy action in Eq. \eqref{GSaction} implies that at long wavelengths, the Goldstone modes have a linear dispersion. This is generally true when the number of Goldstone modes is equal to the number of broken symmetry generators, and results from the fact that $\rho = 0$ excludes terms of the form $\phi_\mu\partial_\tau \phi_\nu$ in the effective Goldstone action \cite{Watanabe2}. 

As a final comment, let us point out that the U$_V(1)$ symmetry is only an emergent symmetry, and is in fact weakly broken in the narrow bands of a complete microscopic model of MATBG. As a result, the Goldstone modes will acquire a tiny mass. We will ignore this mass, however, as it is much smaller than all the other energy scales in the problem. Similarly, the intervalley exchange coupling will introduce a mass for the three Goldstone modes $\phi_i$ of the order of $0.1$ meV. We will also ignore this mass, because it only significantly modifies $K(\q)$ at very small $\q$. These small momenta do not play an important role anyway for our discussion below.

\section{Doping the K-IVC state and electron-Goldstone mode coupling}\label{sec:dopedKIVC}

\begin{figure}
\begin{center}
a)
\includegraphics[scale=0.32]{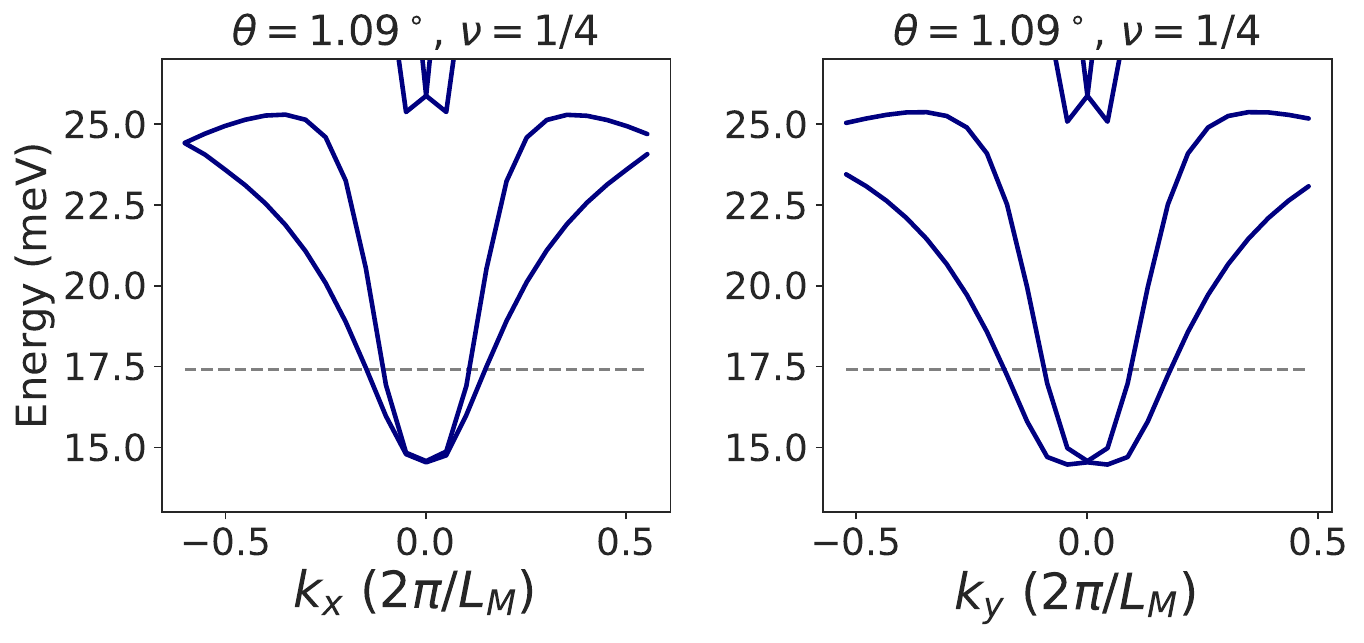}\\
b)
\includegraphics[scale=0.34]{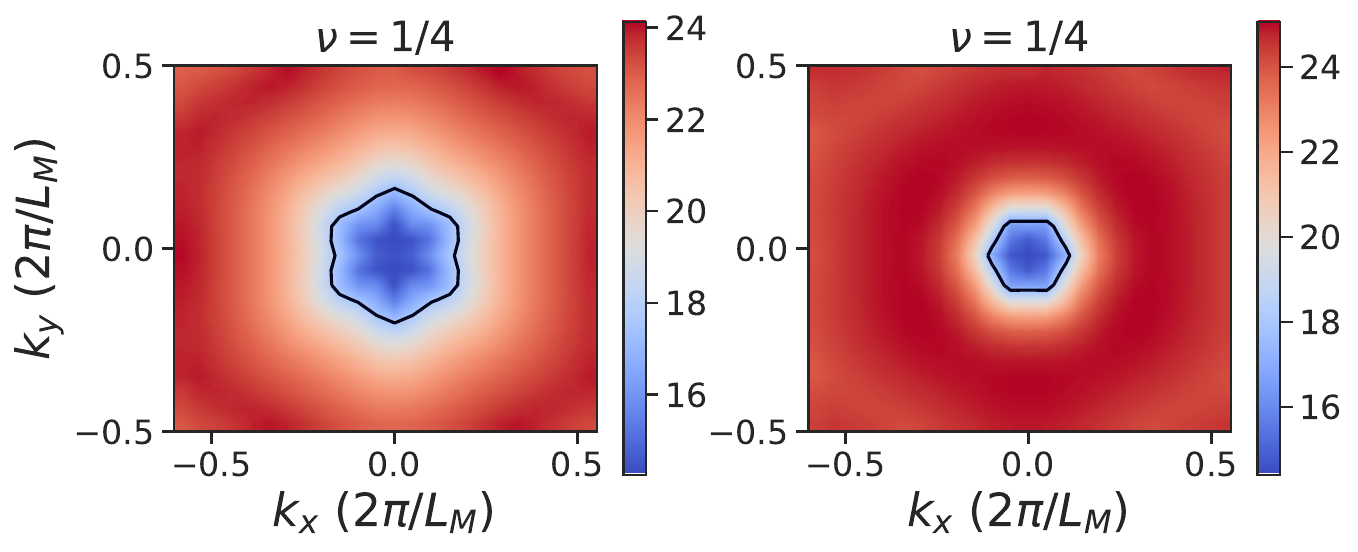}
\caption{(a) K-IVC conduction bands at $\theta=1.09^\circ$ with the Fermi energy at $\nu = 1/4$ indicated by a gray dashed line. (b) Colorplots of the band energies of the two K-IVC conduction bands with a contour corresponding to the two Fermi pockets around $\Gamma$ at $\nu=1/4$.}\label{fig:Ef}
\end{center}
\end{figure}

\begin{figure}
    \centering
    \includegraphics[scale=0.38]{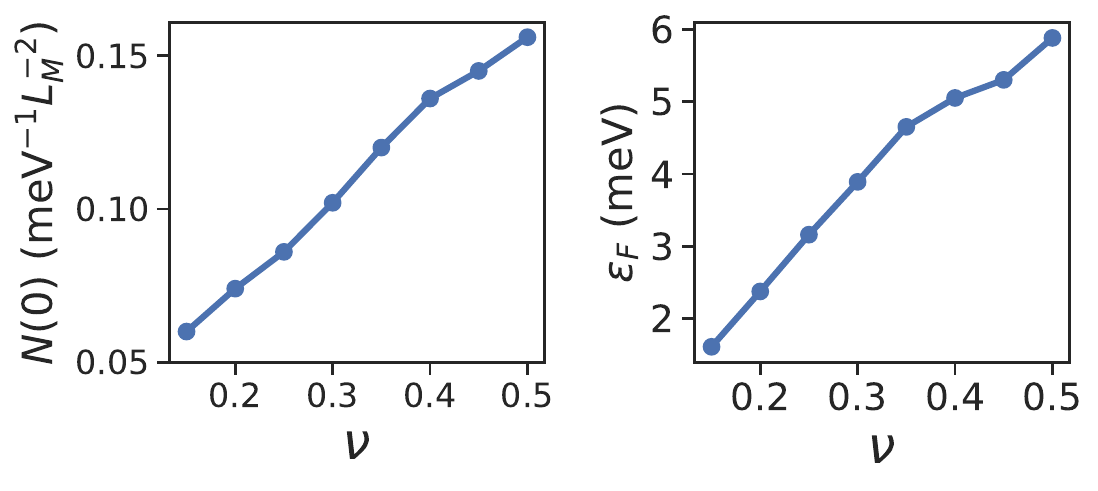}
    \caption{Density of states at the Fermi energy $N(0)$ and the Fermi energy $\varepsilon_F$ (relative to the band minimum) of the K-IVC conduction bands as a function of the filling factor $\nu$.}
    \label{fig:N0Ef}
\end{figure}

In this section, we consider what happens upon doping away from the charge neutrality point. For concreteness, we focus on electron doping and work at a fixed filling $\nu = 1/4$ (i.e., one electron for every four moir\'e unit cells). By solving the Hartree-Fock self-consistency equations at $\nu=1/4$, we find that the K-IVC bands do not change significantly compared to those at charge neutrality. The additional doped electrons simply occupy the lowest-energy states in the conduction bands of the band structure at charge neutrality, without any major changes to the dispersion or the energy gap between valence and conduction bands. In Fig. \ref{fig:Ef}, the K-IVC conduction bands are shown and the Fermi energy at $\nu = 1/4$ is indicated by a gray dashed line. It lies approximately $3.2$ meV above the conduction band minimum. From Fig. \ref{fig:Ef}, we see that at $\nu=1/4$ there are two Fermi surfaces around the $\Gamma$ point. The average Fermi velocity for electrons at the outer (inner) Fermi surface is $v_{F,1} \approx 5.5$ meV$\times L_M$ ($v_{F,2} \approx 8$ meV$\times L_M$). Throughout this work, we will use a notation where subscript $1$ ($2$) refers to the lower (upper) K-IVC conduction band containing the outer (inner) Fermi surface. In Fig.~\ref{fig:N0Ef}, we show the Fermi energy $\varepsilon_F$ and the density of states at the Fermi energy $N(0)$ as a function of the filling $\nu$. We see that for $\nu=1/4$, the density of states is given by $N(0)\approx 0.08$ meV$^{-1}L_M^{-2}$. Note that this value is significantly smaller than the density of states in the BM bands at the magic-angle, where the density of states is $N(0)_{BM}\gtrsim 1$ meV$^{-1}L_M^{-2}$.

We now derive the coupling between the doped electrons in the conduction bands and the K-IVC Goldstone modes. Similarly to the calculation of $K(\q)$ in Appendix~\ref{app:stiffness}, we start from the Hamiltonian \cite{Watanabe} 

\begin{equation}\label{Hphi}
H[\phi_{\mu}] = H_{0} + e^{-i\hat{Q}}H_{\Delta}e^{i\hat{Q}}\,, 
\end{equation}
where 

\begin{eqnarray}
\hat{Q} & = & \frac{1}{2}\int\mathrm{d}\r\; \phi_\mu(\r)f^\dagger_{\r}\tau^zs^\mu f_{\r} 
\end{eqnarray}
is the operator that generates spatially-dependent valley U$(1)$ rotations and opposite spin rotations in the two valleys, with $\phi_\mu(\r)$ being the corresponding Goldstone fields. The operators $f_\r^\dagger$ are the real-space fermion creation operators of the MATBG continuum model~\cite{Morell2010,Santos,BM} in the orbital basis. Going to momentum space and transforming to the band basis of the BM model we can write $\hat{Q}$ as

\begin{eqnarray}
 \hat{Q} & = & \frac{1}{2\sqrt{A}}\sum_{\k,\q}\phi_{\q,\mu} c^\dagger_{\k+\q}\Lambda_\q(\k)\tau^z s^\mu c_\k\,, \label{Q}
\end{eqnarray}
where $A$ is the area of the system and $c^\dagger_{m,\k}$ create electrons in the BM bands, with $m = (\tau,s,a)$ a combined valley, spin, and band index. The sum over $\k$ is restricted to the first mini-BZ, while the sum over $\q$ runs over all BZs in the repeated zone scheme. In Eq.~\eqref{Q}, the form factors $\Lambda_\q(\k)$ result from doing the unitary transformation from the orbital basis to the BM band basis and they are defined as

\begin{equation}\label{formfactor}
    \left[\Lambda_\q(\k)\right]_{mn} = \langle v_{m,\k+\q}|v_{n,\k}\rangle\, ,
\end{equation}
where $|v_{m,\k}\rangle$ are the periodic part of the Bloch states of the BM Hamiltonian. Note that since the BM bands have a well-defined valley and spin quantum number, and do not depend on the spin quantum number, we could put the matrices $\tau^zs^\mu$ outside of the form factor in Eq.~\eqref{Q}. 

Expanding $H[\phi]$ to first order in $\phi$ we find

\begin{eqnarray}\label{expansion}
    H[\phi] 
    & = & H - i[\hat{Q},H_\Delta] \,,
\end{eqnarray}
where $H = \sum_\k c_\k^\dagger[h_0(\k)+\Delta(\k)]c_\k$ is the K-IVC mean-field Hamiltonian. Written out explicitly, the first order term takes the form

\begin{align}
-i[\hat{Q},H_\Delta] & = \frac{i}{2\sqrt{A}} \times\\
\sum_{\k,\q}\phi_{\q,\mu}\, c^\dagger_{\k+\q}&\left[\Delta(\k+\q)\Lambda_\q(\k)\tau^z  - \Lambda_\q(\k)\tau^z\Delta(\k) \right] s^\mu c_\k. \nonumber 
\end{align}
We now perform a transformation to the eigenbasis of the K-IVC Hamiltonian, where $H = \sum_{\alpha,s,\k}E_{\alpha,\k}\psi^\dagger_{\alpha,s,\k}\psi_{\alpha,s,\k}$. The electron operators $\psi^\dagger_{\alpha,s,\k}$, labeled by a K-IVC band index  $\alpha$ and a spin index $s$, are related to the electron operators in the BM basis as

\begin{equation}
    \psi^\dagger_{\alpha,s,\k} = \sum_m \left[u_{\alpha,\k}\right]_m c^\dagger_{m,s,\k}\, ,
\end{equation}
where $\left[u_{\alpha,\k}\right]_m$ are the components of the K-IVC eigenstates $|u_{\alpha,\k}\rangle$. In the K-IVC eigenbasis, the first order term in $\phi$ takes the form

\begin{equation}\label{firstorder}
-i[\hat{Q},H_\Delta] =  \frac{1}{\sqrt{A}}\sum_{\k,\q}\sum_{\alpha,\beta,\nu}g_{\alpha\beta}(\k,\q)\phi_{\q,\mu}\psi^\dagger_{\alpha,\k+\q}s^\mu\psi_{\beta,\k}\, ,
\end{equation}
where the electron-boson coupling $g_{\alpha\beta}(\k,\q)$ is given by

\begin{align}\label{coupling}
g_{\alpha\beta}(\k,\q) &=  \\
\frac{i}{2}\langle u_{\alpha,\k+\q}|(\Delta(\k+\q)\Lambda_\q(\k)\tau^z -& \tau^z\Lambda_\q(\k)\Delta(\k) )|u_{\k,\beta}\rangle. \nonumber
\end{align}
From $\Lambda_{0}(\k)= \mathds{1}$, we see that the coupling at zero momentum transfer can be written as \cite{Watanabe}

\begin{eqnarray}
g_{\alpha\beta}(\k,0) & = & \frac{i}{2}\langle u_{\alpha,\k}|[\Delta(\k),\tau^z]|u_{\beta,\k}\rangle \nonumber \\
 & = &  \frac{i}{2}\langle u_{\alpha,\k}|\tau^z|u_{\beta,\k}\rangle\times \left(E_{\alpha,\k}-E_{\beta,\k}\right)\, ,\label{kzerog}
\end{eqnarray}
where we have used that $[\Delta(\k),\tau^z] = [h_{0}(\k)+\Delta(\k),\tau^z]$. Written in this form, it is clear that the intraband scattering processes vanish at zero momentum transfer. This important property implies that, at small coupling, the Goldstone modes are not Landau damped, and also do not destroy the Landau quasi-particles \cite{Watanabe}.

\section{Goldstone-mediated interaction}\label{sec:SC}

The coupling to the Goldstone modes leads to an effective attractive interaction between the electrons, similar to the familiar attractive interaction in coupled electron-phonon systems. The interaction between electrons doped into the K-IVC insulator generated by the exchange of Goldstone modes is given by

\begin{eqnarray}\label{GSint}
    H_G = \frac{1}{2A}&&\sum_{\q,\k,\k'}V^G_{\alpha\beta\lambda\sigma}(\q,\k,\k')s^\mu_{s_1s'_1}s^\mu_{s_2s'_2} \nonumber\\
    & &\times\psi^\dagger_{\alpha,s_1,\k+\q}\psi^\dagger_{\lambda,s_2,\k'-\q}\psi_{\sigma,s'_2,\k'}\psi_{\beta,s'_1,\k}\, , 
\end{eqnarray}
where the summation over repeated band indices $\alpha,\beta,\lambda,\sigma$, spin indices $s_1,s'_1,s_2,s'_2$ and $\mu = 0,x,y,z$ is implicit. Note that the sum over $\q$ in Eq.~\eqref{GSint} runs over all mini-BZs in the repeated zone scheme. The potential $V^G(\q,\k,\k')$ is given by

\begin{equation}\label{GSpotential}
    V^G_{\alpha\beta\lambda\sigma}(\q,\k,\k') = g_{\alpha\beta}(\k,\q)D(0,\q)g_{\lambda\sigma}(\k',-\q)\, ,
\end{equation} 
where $D(0,\q) = - K(\q)^{-1}$ is the Goldstone mode propagator defined in Eq.~\eqref{GSprop} evaluated at zero frequency and $g_{\alpha\beta}(\k,\q)$ is the coupling function defined in Eq.~\eqref{coupling}.

Note that in the previous section we have calculated $K(\q)$ at charge neutrality, whereas in this section we are considering the K-IVC at non-zero doping, i.e. away from charge neutrality. The fact that the interaction mediated by the Goldstone modes at charge neutrality and at finite doping have approximately the same form, Eq.~\eqref{GSint}, is a non-trivial result. We discuss it in detail in Appendix~\ref{app:lowE}, where we use the path-integral formalism to integrate out the valence band fermionic degrees of freedom and derive the effective low-energy theory that couples Goldstone modes and the conduction band electrons. After carefully summing up certain sets of diagrams we end up with the conclusion that, to good accuracy, the effective interaction between the electrons on the Fermi surface mediated by Goldstone modes is given by the same expression that one would obtain by integrating out Goldstone modes at charge neutrality, i.e., Eqs.~\eqref{GSint}-\eqref{GSpotential}. We emphasize that this interaction should not be viewed as some low-energy starting point that needs to be further renormalized by, e.g., particle-hole modes. Instead, it is an effective interaction that already takes into account important renormalization effects and will be used directly to calculate superconducting instabilities. This is somewhat similar in spirit to the Eliashberg theory of superconductivity, where one self-consistently solves for the electron Green's function, while taking electron-phonon interaction as an input parameter (we, however, do not study the frequency dependence of the gap function in this paper, but do a BCS-type analysis instead). 

As a preparatory step for studying superconductivity, we decompose the Goldstone mediated interaction in the different Cooper channels. In particular, we use the following two Fierz identities:

\begin{equation}
\delta_{s_1s'_1}\delta_{s_2s'_2} = \frac{1}{2}s^y_{s_1s_2}s^y_{s'_2s'_1} + \frac{1}{2} (s^ys^i)_{s_1s_2}(s^is^y)_{s'_2s'_1}
\end{equation}
\begin{equation}
s^j_{s_1s'_1}s^j_{s_2s'_2} =  -\frac{3}{2}s^y_{s_1s_2}s^y_{s'_2s'_1} + \frac{1}{2} (s^ys^i)_{s_1s_2}(s^is^y)_{s'_2s'_1}\, ,
\end{equation}
(with implicit summation over the repeated indices $i,j=x,y,z$) to rewrite the effective interaction as

\begin{widetext}
\begin{equation}
    H_G = \frac{1}{2A}\sum_{\q,\k,\k'}2V^{G}_{\alpha\beta\lambda\sigma}(\q,\k,\k')\left[
    -\frac{1}{2}(\psi^\dagger_{\alpha\k+\q}s^y\psi^\dagger_{\lambda,\k'-\q})(\psi_{\sigma,\k'}s^y\psi_{\beta,\k}) + \frac{1}{2}(\psi^\dagger_{\alpha\k+\q}s^ys^j\psi^\dagger_{\lambda,\k'-\q})(\psi_{\sigma,\k'}s^js^y\psi_{\beta,\k})\right]\, , 
\end{equation}
\end{widetext}
Note the minus sign in front of the spin-singlet part of the interaction. This implies that the Goldstone-mediated interaction in the spin-singlet channel is actually repulsive instead of attractive. Physically, this can be understood from the fact that the combined exchange of the three Goldstone modes corresponding to opposite spin rotations in the two valleys generates an effective ferromagnetic interaction, which favors pairing in a spin-triplet channel. This is completely analogous to how ferromagnetic fluctuations drive spin-triplet superconductivity in He$^3$ \cite{Leggett}. It is also similar to how \emph{anti-ferromagnetic} fluctuations in doped Mott insulators favor pairing in \emph{spin-singlet} channels \cite{Scalapino1,Scalapino2}.

\section{Superconducting instabilities}\label{sec:instabilities}

We focus on zero-momentum, spin-triplet pairing of electrons near the Fermi surfaces of the doped K-IVC state. The part of the Goldstone-mediated interaction potential that we are interested in is therefore given by

\begin{equation}\label{tripletV}
    \hat{V}^G_{\alpha\beta}(\k',\k) = \sum_\G 2V^G_{\alpha\beta\alpha\beta}(\k'-\k+\G,\k,-\k)\, ,
\end{equation}
where the sum over moir\'e reciprocal lattice vectors $\G$ takes into account the umklapp processes due to the superlattice potential. In our numerical calculations, we restrict $\G$ to lie within the first three shells of mini-BZ. This part of the interaction scatters a spin-triplet Cooper pair $c^\dagger_{\k,\alpha}is^ys^jc^\dagger_{-\k,\alpha}$ in band $\alpha$ to a spin-triplet Cooper pair $c^\dagger_{\k,\beta}is^ys^j c^\dagger_{-\k,\beta}$ in band $\beta$. The factor of $2$ in Eq.~\eqref{tripletV} comes from the fact that both the valley-U$(1)$ Goldstone mode and the SU$(2)$ Goldstone modes from opposite spin rotations in the two valleys contribute equally to the triplet channel, as explained in the previous section.  

In order to obtain physically relevant results for the superconducting instabilities, it is important not to ignore the repulsive Coulomb interaction. The metallic state at non-zero doping will screen the Coulomb interaction. The part of the resulting screened Coulomb potential which scatters zero momentum Cooper pairs in the different K-IVC bands is given by 

\begin{equation}
\hat{V}^C_{\alpha\beta}(\k',\k) = \sum_\G V_{\alpha\beta\alpha\beta}^{C,scr}(\k'-\k+\G,\k,-\k)\, ,
\end{equation}
where the screened Coulomb potential $V^{C,scr}_{\alpha\beta\lambda\sigma}(\q,\k,\k')$ is discussed in more detail in Appendix \ref{app:repulsive}.

Because of the Kramers time-reversal symmetry of the K-IVC state, the Goldstone-mediated and Coulomb interaction potentials can be written as 
 
\begin{widetext}

\begin{eqnarray}
\hat{V}^G_{\alpha\beta}(\k',\k) & = & e^{i\varphi_\alpha(\k')}\left(\sum_\G D(0,\k'-\k+\G)  |g_{\alpha\beta}(\k,\k'-\k+\G)|^2 \right)e^{-i\varphi_\beta(\k)}\, ,\label{positive2} \\
\hat{V}^C_{\alpha\beta}(\k',\k) & = & e^{i\varphi_\alpha(\k')}\left(\sum_\G \frac{V_C(|\k'-\k+\G|)}{\epsilon(\k'-\k+\G)} \big| \left[ F_{\k'-\k+\G}(\k)\right]_{\alpha\beta}\big|^2\right)e^{-i\varphi_\beta(\k)}, \label{positive1}
\end{eqnarray}
\end{widetext}
where $e^{i\varphi_\alpha(\k')}$ and $e^{-i\varphi_\beta(\k)}$ are gauge-dependent phase factors (see Appendix~\ref{app:Kramers}). These phases are irrelevant for our analysis of the superconducting instabilities and we can simply omit them for now, which makes the interaction potentials real-valued. At the end of this section, we will reintroduce the gauge-dependent phase factors. As explained in Appendix \ref{app:repulsive}, the factor $\epsilon(\q)^{-1}$ in Eq. \eqref{positive1} incorporates the screening of the Coulomb interaction by the doped electrons.   

Next, we define the total interaction potential in the Cooper channel for electrons on the Fermi surface as 

\begin{align}
    V_{\alpha\beta}(\theta',\theta) &= \hat{V}^C_{\alpha\beta}[\k_{F,\alpha}(\theta'),\k_{F,\beta}(\theta)] \nonumber \\ &+ \hat{V}^G_{\alpha\beta}[\k_{F,\alpha}(\theta'),\k_{F,\beta}(\theta)]\, ,
\end{align}
where $\theta$ and $\theta'$ are polar angles in momentum space and $\k_{F,\alpha}(\theta)$ is the (angle-dependent) Fermi momentum on the Fermi surface of band $\alpha$.

As discussed in the previous section, the Goldstone-mediated interaction is repulsive in the spin-singlet channels. Inclusion of the Coulomb interaction  leads to an instability in the $d$-wave pairing channel through the Kohn-Luttinger mechanism~\cite{KohnLuttinger1965,MaitiChubukov2013}. However, this instability is subleading and has nothing to do with the Goldstone modes, so we will  exclusively focus on spin-triplet channels from now on.  
We will thus look for superconducting instabilities with gap functions of the form

\begin{equation}
    \tilde{{\boldsymbol \Delta}}_\k = \left(\begin{matrix}\tilde{ \Delta}_{1,\k} & 0\\0 & \tilde{ \Delta}_{2,\k} \end{matrix}\right)\otimes is_y{\boldsymbol s}\, , \label{gaptriplet}
\end{equation}
where $\tilde{\Delta}_{\alpha,\k}$ is the gap function in the band labeled by $\alpha$. Note that we use a tilde to distinguish the superconducting gap from the K-IVC order parameter. 

To find superconducting instabilities we can solve the linearized gap equation equation~\footnote{Note that this equation is approximate since it relies on the assumption of an isotropic spectrum on the Fermi surface.}

\begin{equation}\label{eval}
    \sum_\beta \int\frac{\mathrm{d}\theta}{2\pi} V_{\alpha\beta}(\theta',\theta) N_\beta(0) \tilde{\Delta}_\beta(\theta) = -\lambda \tilde{\Delta}_\alpha(\theta')\, ,
\end{equation}
where $\lambda>0$ and $N_\beta(0)$ is the density of states per spin at Fermi surface $\beta$, given by

\begin{equation}
    N_\beta(0) = \int \frac{\mathrm{d}\theta}{2\pi} \frac{k_{F,\beta}(\theta)}{2\pi}\left| \frac{\partial E_{\beta}(k,\theta)}{\partial k} \right|^{-1}_{k = k_{F,\beta}(\theta)}.
\end{equation}
Note that $N_1(0)+N_2(0)= N(0)/2$, since we previously defined $N(0)$ to contain a spin degeneracy factor. See Appendix~\ref{app:BCS} for a review on the derivation of the linearized gap equation.

To find the solutions to Eq.~\eqref{eval}, we go to the angular momentum basis and define

\begin{equation}
    V_{\alpha\beta}(m,n) = \int\frac{\mathrm{d}\theta'}{2\pi}\int\frac{\mathrm{d}\theta}{2\pi} e^{i m \theta'}V_{\alpha\beta}(\theta',\theta)e^{-i n\theta}.
\end{equation}
Because $V_{\alpha\beta}(m,n)$ is real (recall that we ignore the gauge-dependent phase factors in Eqs.~\eqref{positive1} and~\eqref{positive2} for now), it follows that $V^*_{\alpha\beta}(m,n) = V_{\alpha\beta}(-m,-n)$. Also, because the K-IVC state is invariant under the mirror symmetry $(x,y) \rightarrow (x,-y)$, it follows that $V_{\alpha\beta}(-\theta',-\theta)=V_{\alpha\beta}(\theta',\theta)$. This implies that $V_{\alpha\beta}(m,n) = V_{\alpha\beta}(-m,-n) = V^{*}_{\alpha\beta}(m,n)$, such that that the components $V_{\alpha \beta}(m,n)$ are real. From the six-fold in-plane rotation symmetry of the K-IVC state, it also follows that $V_{\alpha\beta}(m,n) = 0$ if $m\neq n$ mod $6$. Because we are considering spin-triplet pairing only odd angular momenta contribute, which means that we restrict $m$ and $n$ in $V_{\alpha\beta}(m,n)$ to both be odd.

For the numerical calculations, we further restrict ourselves to angular harmonics $e^{in\theta}$ with $|n|\leq 6$. This converts the eigenvalue equation~\eqref{eval} to a set of decoupled finite-dimensional matrix eigenvalue equations
 
\begin{equation}\label{gapn}
\sum_{\beta,M} V_{\alpha\beta}(n+6N,n+6M) N_{\beta}(0) \tilde{\Delta}^n_{\beta,M} = -\lambda_n \tilde{\Delta}^n_{\alpha,N}\,,
\end{equation}
labeled by $n\in\{\pm 1, 3\}$. In Eq.~\eqref{gapn}, we have defined $\tilde{\Delta}_{\alpha,N}^n \equiv \tilde{\Delta}_{\alpha,n+6N}$, and $\tilde{\Delta}_{\alpha,n} = \int\frac{\mathrm{d}\theta}{2\pi} e^{in\theta}\tilde{\Delta}_\alpha(\theta)$. Numerically, the summation over $M$ is restricted by the requirement that $|n+6M|\leq 6$ (same holds for $N$, i.e., $|n+6N|\leq 6$). Solutions to the linearized gap equation with $n=\pm 1$ correspond to pairing in the degenerate $p$-wave channels, and $n = 3$ to pairing in the $f$-wave channel.

The gap functions on the Fermi surfaces corresponding to the different $\lambda_n$ are given by the eigenvectors $\tilde{\Delta}_{\alpha,N}^n$:

\begin{equation}
    \tilde{\Delta}^n_{\alpha}(\theta) = \sum_{N}\tilde{\Delta}^n_{\alpha,N}e^{-i(n+6N)\theta}.
\end{equation}
At this point, it is straightforward to reintroduce the gauge-dependent phase factors $e^{i\varphi_\alpha(\theta)}$. These phases do not change the values of $\lambda_n$, but only modify the gap function to take the form

\begin{equation}
    \tilde{\Delta}^n_{\alpha}(\theta) = e^{i\varphi_\alpha(\theta)} \sum_{N}\tilde{\Delta}^n_{\alpha,N}e^{-i(n+6N)\theta}
\end{equation}

\begin{figure}
    \centering
    \includegraphics[scale=0.4]{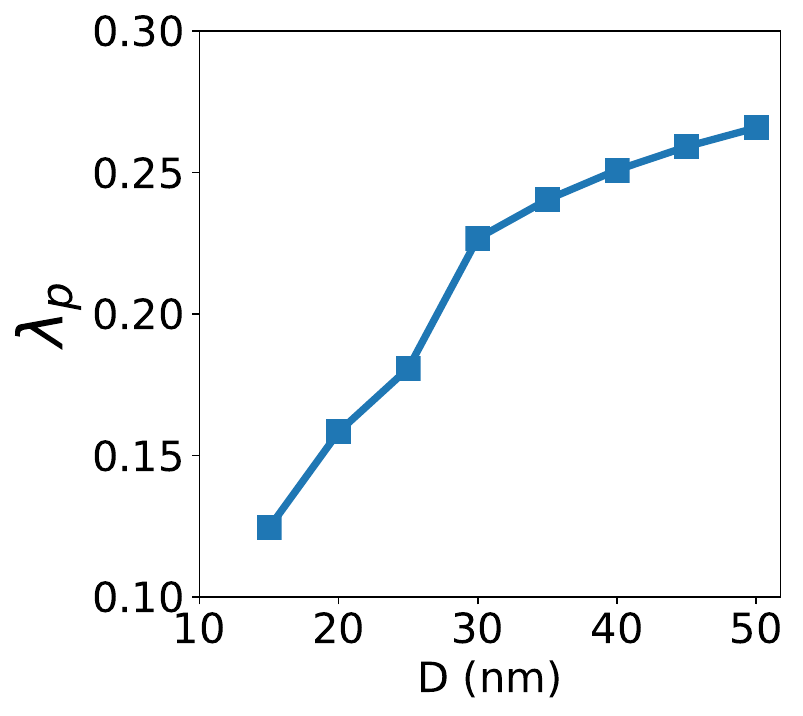}
    \caption{Dimensionless pairing strength of the superconducting $p-$wave channel as a function of the gate distance $D$.}
    \label{fig:lambdas}
\end{figure}
Solving the linearized gap equation  for spin-triplet pairing numerically, we find that there is only a superconducting instability in the $p$-wave channel. In Fig.~\ref{fig:lambdas}, we plot the numerically obtained value for $\lambda_p$ as a function of gate distance $D$. These results were obtained using $N_1(0) = 2.9\times 10^{-2}$ meV$^{-1}$ $L_M^{-2}$ and $N_2(0) = 1.1\times 10^{-2}$ meV$^{-1}$ $L_M^{-2}$. We see from Fig.~\ref{fig:lambdas} that the $p$-wave pairing strength increases as a function of gate distance. In the experiments of Ref.~\cite{EfetovScreening}, the gate distance of the device displaying superconductivity around charge neutrality was $~ 15$ nm, for which we find $\lambda \sim 0.12$. Using the standard BCS analysis (see Appendix~\ref{app:BCS}), one obtains an estimate for the critical temperature given by

\begin{equation}
    k_BT_c \sim \varepsilon_F\times e^{-1/\lambda}\, .
\end{equation}
Using $\varepsilon_F = 3.2$ meV and $\lambda = 0.12$, we obtain $T_c \sim 8.94 \times 10^{-3}$ K. This value for $T_c$ is much smaller than the experimental value of $T_c \approx 0.3$ K \cite{EfetovScreening}. Nevertheless, given that we have taken the effect of the repulsive Coulomb interaction into account in our calculation, the fact that there is a pairing instability at the BCS level is a non-trivial result. We expect that our estimate for $T_c$ is too small, and that a more sophisticated calculation (such as for example a renormalization group calculation where the net attractive interaction can grow under scaling towards lower energies), will lead to a significantly higher value for $T_c$. Our naive BCS calculation already shows that it is worthwhile to apply more advanced techniques to the problem of electrons coupled to the collective modes of a doped K-IVC insulator. In particular, we expect that allowing for a general frequency dependence of the gap function should significantly improve the stability of the superconductor.

\section{Discussion}\label{sec:discussion}

To summarize, we have studied the connection between the insulating state observed in MATBG at charge neutrality~\cite{efetov,EfetovScreening,Wong}, and the neighboring superconducting domes~\cite{efetov,EfetovScreening}. Our starting point was to identify the insulating state with the K-IVC insulator of Ref.~\cite{KIVC}, which spontaneously breaks the valley-U$(1)$ symmetry and the SU$(2)$ symmetry corresponding to opposite spin rotations in the two valleys. We have coupled the charge carriers at finite doping $(\nu \sim 0.25)$ to the four Goldstone modes of the K-IVC state. The resulting effective interaction mediated by the exchange of these Goldstone modes is attractive in the spin-triplet channels, but repulsive in the singlet channels. This gives rise to a general mechanism for obtaining spin-triplet superconductivity from an inter-valley coherent normal state in MATBG. A na\"{i}ve BCS calculation which incorporates both the effective interaction from Goldstone-mode exchange and the repulsive Coulomb interaction shows that there is indeed a pairing instability of the doped K-IVC state. We find that the leading pairing instability occurs in the $p-$wave channel.

There is room for improvement of the results presented here. Firstly, our starting point is a self-consistent Hartree-Fock band spectrum, and therefore ignores many-body correlation effects. Secondly, our analysis of the interactions that lead to the superconducting instability relies on an RPA approximation, and thus ignores the effects of quantum fluctuations. Thirdly, our numerics are done on a $24\times 24$ momentum grid, which introduces unknown finite-size errors. And finally, the critical temperature obtained via the standard BCS formula is smaller than the experimental value~\cite{efetov,EfetovScreening} by a factor of $\sim 30$ (due to the exponential sensitivity of $T_c$ on $\lambda$, this would correspond to a value for $\lambda$ that is less than two times larger than found here). However, we hope that the approach put forward in this paper can be used as a starting point for future theoretical analytical and numerical works. In particular, we believe that a na\"{i}ve BCS analysis slightly underestimates $\lambda$, and more advanced approaches can potentially give higher critical temperatures.

The present work opens up several other questions which deserve further investigation. For example, it would be interesting to understand whether the pairing mechanism discussed in this work could also be operative near $\nu = \pm 2$, where a spin polarized K-IVC state is realized \cite{KIVC}.

The Hamiltonian of MATBG restricted to the nearly flat bands has an approximate $U(4)\times U(4)$ symmetry~\cite{KIVC,KangVafekPRL}, which is responsible for the close competition between many different symmetry-broken phases in self-consistent Hartree-Fock studies \cite{XieMacDonald,CaltechSTM,LiuKhalaf,KIVC,CeaGuinea}. Because of this approximate symmetry, we expect  the existence of many nearly soft bosonic modes corresponding to fluctuations within the low-energy $U(4)\times U(4)$-manifold. In principle, all these modes can be important for superconductivity. One of these modes even becomes massless at the continuous phase transition between the K-IVC state and the valley-Hall state, which occurs when the microscopic sublattice splitting induced by the hexagonal boron nitride substrate is around $~ 10$ meV \cite{KIVC}. In Appendix~\ref{app:VH}, we argue that non-Fermi liquid physics is expected at this critical point, but that deviations from Fermi liquid theory will only manifest themselves on very long distance and time scales. An interesting topic for future work is to further study the role of these nearly soft or critical bosonic modes.

\section{Acknowledgments}
M. Z. and N. B. would like to thank Ashvin Vishwanath, Eslam Khalaf, Shubhayu Chatterjee, and Shang Liu for collaboration on related projects, which were vital for the results obtained in this work. We also want to thank Jonathan Ruhman, Ashvin Vishwanath, Eslam Khalaf, and Shubhayu Chatterjee for useful discussions and comments. V. K. was supported by the Quantum Materials program at LBNL, funded by the US Department of Energy under Contract No. DE-AC02-05CH11231. M. Z. was supported by the Director, Office of Science, Office of Basic Energy Sciences, Materials Sciences and Engineering Division of the U.S. Department of Energy under contract no. DE-AC02-05-CH11231 (van der Waals heterostructures program, KCWF16).

\bibliography{bib}

\begin{appendix}
\onecolumngrid


\section{Calculation of $K(\q)$ and $\rho_s$ at charge neutrality}\label{app:stiffness}

In this Appendix, we demonstrate how to calculate function $K(\bq)$ and parameter $\rho_s$ introduced in Eq.~\eqref{GSprop} at charge neutrality, i.e., at zero doping. We start by defining the operator which generates spatially dependent valley U$(1)$ rotations and opposite spin rotations in the two valleys:
\begin{equation}
\hat{Q} = \frac{1}{2}\int\mathrm{d}\br\; \phi_\mu(\br)f^\dagger_\br \tau^z s^\mu f_\br \, ,
\end{equation}
and write it in the band basis of the BM model to obtain
\begin{equation}
\hat{Q} = \frac{1}{2\sqrt{A}}\sum_{\bk,\bq} \phi_{\bq,\mu} c^\dagger_{\bk+\bq}\Lambda_{\bq}(\bk)\tau^zs^\mu c_{\bk}\, ,
\end{equation}
where $A$ is the area of the system and $\Lambda_{\bq}(\bk)$ are the form factors defined in Eq.~\eqref{formfactor}. Next, we choose a particular gauge for the K-IVC order parameter and define the following Hamiltonian:

\begin{equation}\label{hamphi}
H[\phi] = H_{0} + e^{-i\hat{Q}}H_{\Delta}e^{i\hat{Q}}\, ,
\end{equation}
where, in accordance with the main text, $H_{0}$ is the valley symmetric part of the K-IVC Hamiltonian and $H_{\Delta}$ is the K-IVC order parameter, $H_{\Delta} = \sum_{\bk}c^\dagger_\bk \Delta(\bk)c_\bk$.

Using the Hamiltonian in Eq.~\eqref{hamphi} we obtain the free energy $F[\phi]$ as

\begin{equation}
\mathcal{Z}[\phi(\br)]= e^{-\beta F[\phi(\br)]} = \text{Tr}\left(e^{-\beta H[\phi(\br)]} \right),
\end{equation}
where the trace is over fermionic degrees of freedom. We can now do an expansion of the free energy in $\phi$ and write

\begin{eqnarray}
F[\phi(\br)] & = & F_0 + \frac{1}{2}\int\mathrm{d}^2\br\, \phi_\mu(\r)\hat{K}_{\mu\nu} \phi_\nu(\r) + \dots \, ,
\end{eqnarray}
where $\hat{K}$ is a general differential operator. Going to momentum space, we obtain

\begin{eqnarray}
F[\phi(\q)] & = & F_0 + \frac{1}{2}\sum_{\q}\, \phi_\mu(\q)K_{\mu\nu}(\q)\phi_\nu(-\q) + \dots\, ,
\end{eqnarray}
In our discussion in the main text, we are interested in the function $K_{\mu\nu}(\q)$, which we can obtain as

\begin{equation}\label{def}
K_{\mu\nu}(\q) =  \frac{\delta^2 F[\phi]}{\delta\phi_\mu(\bq)\delta\phi_\nu(-\bq)}\Bigg|_{\phi = 0} = -\frac{1}{\beta}  \frac{\delta^2 \ln\mathcal{Z}[\phi]}{\delta\phi_\mu(\bq)\delta\phi_\nu(-\bq)}\Bigg|_{\phi = 0} \, .
\end{equation}
To obtain a more convenient formula for $K_{\mu\nu}(\q)$ from Eq.~\eqref{def}, we first expand the Hamiltonian to second order in $\phi_\mu$, which gives

\begin{equation}
H[\phi]  =  H_{0} + H_K -i[\hat{Q},H_\Delta] -\frac{1}{2}[\hat{Q},[\hat{Q},H_\Delta]]+ \dots
\end{equation}
Writing this in the K-IVC band basis, we obtain 

\begin{eqnarray}
H[\phi] & = &   \sum_\bk \sum_{\alpha} E_{\alpha,\bk}\psi^\dagger_{\alpha,\bk}\psi_{\alpha,\bk} + \frac{1}{\sqrt{A}}\sum_{\bk,\bq}\sum_{\alpha,\beta} \sum_{\mu}\phi_\mu(\bq) g_{\alpha\beta}(\bk,\bq)\psi^\dagger_{\alpha,\bk+\bq}s^\mu\psi_{\beta,\bk} \\
 & &  + \frac{1}{2 A}\sum_{\bk,\bq,\bq'}\sum_{\alpha,\beta}\sum_{\mu,\nu}\phi_\mu(\bq)\phi_\nu(\bq')\psi^\dagger_{\alpha,\bk+\bq+\bq'}\left[\delta_{\mu\nu}\tilde{g}_{\alpha\beta}(\bk,\bq,\bq') + (1-\delta_{\mu\nu})\bar{g}_{\alpha\beta}^{\mu\nu}(\bk,\bq,\bq') s^\mu s^\nu \right]\psi_{\beta,\bk}\, , \nonumber
\end{eqnarray}
where
\begin{eqnarray}
g_{\alpha\beta}(\bk,\bq) & = & \frac{i}{2}\langle u_{\alpha,\bk+\bq}|\left(\Delta(\bk+\bq)\Lambda_{\bq}(\bk)\tau^z - \tau^z\Lambda_{\bq}(\bk)\Delta(\bk) \right)|u_{\beta,\bk}\rangle \\
\tilde{g}_{\alpha\beta}(\bk,\bq,\bq') & = & -\frac{1}{4}\langle u_{\alpha,\bk+\bq+\bq'}|\bigg(\Delta(\bk+\bq+\bq')\Lambda_{\bq}(\bk+\bq')\Lambda_{\bq'}(\bk) + \Lambda_{\bq'}(\bk+\bq)\Delta(\bk+\bq)\Lambda_{\bq}(\bk) \nonumber\\ 
 & & +\Lambda_{\bq}(\bk+\bq')\Delta(\bk+\bq')\Lambda_{\bq'}(\bk) + \Lambda_{\bq'}(\bk+\bq)\Lambda_{\bq}(\bk)\Delta(\bk)  \bigg)|u_{\beta,\bk}\rangle\, ,\label{gtilde} 
\end{eqnarray}
and $|u_{\alpha,\bk}\rangle$ are the eigenstates of the K-IVC Hamiltonian. For the derivation of $K_{\mu\nu}(\q)$, we do not need an explicit expression for $\bar{g}^{\mu\nu}_{\alpha\beta}(\k,\q,\q')$. As a next step we write the partition function as a path integral

\begin{equation}
\mathcal{Z}[\phi] = \int D\bar{\psi}D\psi e^{-S}\, ,
\end{equation}
where the imaginary-time action is given by

\begin{eqnarray}
S & = & \sum_{k = \bk,i\omega_n} \sum_{\alpha}\bar{\psi}_\alpha (k)\left(-i\omega_n + E_{\bk,\alpha} \right)\psi_\alpha(k) + \frac{1}{\sqrt{A}}\sum_{k,q} \sum_{\alpha, \beta, \mu} \phi_\mu (q)g_{\alpha\beta}(\bk,\bq)\bar{\psi}_\alpha (k+q) s^\mu \psi_\beta (k) \\
& & +\frac{1}{2A}\sum_{k,q,q'} \sum_{\alpha, \beta, \mu, \nu}\phi_\mu (q)\phi_\nu (q')\bar{\psi}_\alpha(k+q+q')\left[\delta_{\mu\nu}\tilde{g}_{\alpha\beta}(\bk,\bq,\bq') + (1-\delta_{\mu\nu})\bar{g}_{\alpha\beta}^{\mu\nu}(\bk,\bq,\bq') s^\mu s^\nu \right]\psi_\beta(k). \nonumber
\end{eqnarray}
The Feynman rules for the vertices in this action are

\begin{equation}
\begin{tikzpicture}
\begin{feynman}
	\vertex (v1) at (-0.5,0) {\(\bq,\mu\)} ;
	\vertex (v2) at (1,0);
	\vertex (v3) at (2,1) {\(\alpha,s,\,k+q\)};
	\vertex (v4) at (2,-1) {\(\beta,s',\,k\)};
	\vertex (v5) at (4.8,-0.1) {\(=ig_{\alpha\beta}(\bk,\bq)\left[s^\mu \right]_{ss'}\)};
	\diagram*{
		(v1) --[photon] (v2),
		(v2) --[fermion] (v3),
		(v4) --[fermion] (v2)
	}; 
\end{feynman}
\end{tikzpicture}
\end{equation}
and

\begin{equation}
\begin{tikzpicture}
\begin{feynman}
	\vertex (v1) at (0,1) {\(\bq',\nu\)} ;
	\vertex (v2) at (0,-1) {\(\bq,\mu\)};
	\vertex (v3) at (1,0) ;
	\vertex (v4) at (2,1) {\(\alpha,s,\,k+q+q'\)};
	\vertex (v5) at (2,-1) {\(\beta,s',\, k\)};
	\vertex (v6) at (8,-0.1) {\(=-\frac{1}{2}\left[\delta_{\mu\nu}\tilde{g}_{\alpha\beta}(\bk,\bq,\bq')\delta_{ss'} + (1-\delta_{\mu\nu})\bar{g}^{\mu\nu}_{\alpha\beta}(\k,\q,\q')\left[s^\mu s^\nu \right]_{ss'}\right].\)};
	\diagram*{
		(v1) --[photon] (v3),
		(v2) --[photon] (v3),
		(v3) --[fermion] (v4),
		(v5) --[fermion] (v3)
	}; 
\end{feynman}
\end{tikzpicture}
\end{equation}
Using Eq. \eqref{def}, we find that $K_{\mu\nu}(\q)$ is given by the sum of the following two diagrams:

\begin{equation}
\begin{tikzpicture}
\begin{feynman}
	\vertex (v1) at (0,-0.2) {\(\bq\)};
	\vertex (v2) at (1,0.1);
	\vertex (v8) at (1,1);
	\vertex (v3) at (2,-0.2)  {\(-\bq\)};
	\vertex (v4) at (4,0)  {\(\bq\)};
	\vertex (v5) at (5,0);
	\vertex (v6) at (6,0);
	\vertex (v7) at (7,0)  {\(-\bq\)};
	\diagram*{
	(v1) --[photon] (v2),
	(v2) --[fermion,half left] (v8),
	(v8) --[fermion,half left] (v2),
	(v2) --[photon] (v3);
	(v4) --[photon] (v5);
	(v5) --[fermion,half left](v6);
	(v6) --[fermion,half left] (v5);
	(v6) --[photon] (v7);
	};
\end{feynman}
\end{tikzpicture}\, ,
\end{equation}
where the first diagram represents the diamagnetic contribution to the stiffness and the second diagram is the paramagnetic contribution.

The diamagnetic contribution is evaluated to give

\begin{equation}
K^{dia}_{\mu\nu}(\bq) = \delta_{\mu\nu}\frac{2}{A} \sum_{\bk}\sum_\alpha f_{\alpha,\bk} \tilde{g}_{\alpha\alpha}(\bk,\bq,-\bq)\, ,
\end{equation}
where $f_{\alpha,\bk}=f(E_{\alpha,\bk})$ is the Fermi-Dirac distribution function, and the Kronecker-delta $\delta_{\mu\nu}$ and the factor of two come from the trace over spin indices. 

The paramagnetic contribution equals

\begin{equation}
K^{para}_{\mu\nu}(\bq) = \delta_{\mu\nu}\frac{2}{A}\sum_{\bk} \sum_{\alpha,\beta} \frac{f_{\beta,\bk}-f_{\alpha,\bk+\bq}}{E_{\beta,\bk}-E_{\alpha,\bk+\bq}} |g_{\alpha\beta}(\bk,\bq)|^2\, ,
\end{equation}
where the factor $2\delta_{\mu\nu}$ again comes from the trace over spin indices.

Both $K^{dia}_{\mu\nu}(\bq)$ and $K^{para}_{\mu\nu}(\bq)$ are easily evaluated numerically from the mean-field K-IVC band structure, and $K_{\mu\nu}(\q)$ is simply given by $K_{\mu\nu}(\q)=K^{dia}_{\mu\nu}(\q)+K^{para}_{\mu\nu}(\q) = K(\q)\delta_{\mu\nu}$. 

From $K(\q)$, one obtains the K-IVC stiffness $\rho_s$ by fitting (at zero temperature) to the long wavelength part of $K(\bq)= \rho_s\bq^2 + \mathcal{O}(\bq^4)$. Note that the zeroth order term in the long wavelength expansion of $K(\q)$ has to vanish because $\phi$ is a Goldstone mode. To see this explicitly, we write the diamagnetic contribution to $K_{\mu\nu}(0)$ in the original BM band basis, where it is given by

\begin{align}
&-\frac{2\delta_{\mu\nu}}{4\beta A}\sum_{\bk}\sum_{\omega_n}\text{tr}\left([\tau^z,[\tau^z,\Delta(\bk)]]\frac{1}{i\omega_n-h_{0}(\bk)-\Delta(\bk)} \right)  \\
& = \frac{2\delta_{\mu\nu}}{4\beta A}\sum_{\bk}\sum_{\omega_n}\text{tr}\left([\tau^z,\Delta(\bk)]\frac{1}{i\omega_n-h_{0}(\bk)-\Delta(\bk)}[\tau^z,\Delta(\bk)]\frac{1}{i\omega_n-h_{0}(\bk)-\Delta(\bk)} \right)\, ,\label{d1}
\end{align}
where the trace is over the valley and BM band indices. The trace over the spin index has already been performed. To obtain the equality in the second line we have used the identities $\text{tr}([A,B]C) = -\text{tr}(B[A,C])$ and $[A,B^{-1}] = - B^{-1}[A,B]B^{-1}$, where $A, B,$ and $C$ are arbitrary matrices (and $B$ is invertible for the second identity). We have also used that $[\tau^z,h_0(\k)] = 0$. The paramagnetic contribution to $K_{\mu\nu}(0)$ takes the following form in the BM band basis: 

\begin{equation}\label{diagram2}
-\frac{2\delta_{\mu\nu}}{4\beta A} \sum_{\bk}\sum_{\omega_n}\text{tr}\left([\tau^z,\Delta(\bk)]\frac{1}{i\omega_n -h_{0}(\bk)-\Delta(\bk)}[\tau^z,\Delta(\bk)]\frac{1}{i\omega_n -h_{0}(\bk)-\Delta(\bk)} \right).
\end{equation}
By comparing Eq. \eqref{d1} to Eq. \eqref{diagram2}, we see that the diamagnetic and paramagnetic contributions cancel exactly, such that $K(0)=0$ as required.

\section{Low-energy effective theory at non-zero doping}\label{app:lowE}

At non-zero doping, we can write down an effective low-energy theory to describe the electrons at the Fermi surface and the Goldstone modes. The total imaginary time action of this effective theory, obtained by integrating out the electons in the K-IVC valence bands, is a sum of several terms: 

\begin{equation}\label{actions}
S = S_\psi + S_C + S_{\psi-\phi}  + S_{\psi-\phi^2} +  S_\phi \, .
\end{equation}
Below, we define and discuss each of these terms one by one. 

The first term $S_\psi$ is given by
\begin{equation}
S_\psi  = \int\frac{d\omega}{2\pi}  \int_\k \bar{\psi}_\alpha(i\omega,\k)(-i\omega+ E_{\alpha,\k} )\psi_\alpha(i\omega,\k) 
\end{equation}
and describes the electrons in the K-IVC conduction bands, meaning that $\alpha > 0$. Note that spin indices are always implicit, and that we have introduced the notation $\int_\k = \int \frac{\mathrm{d}\k}{(2\pi)^2}$. We will represent propagators of the conduction band electrons diagrammatically in the conventional way, i.e., by a solid straight line with an arrow.

The second term $S_C$ is the Coulomb interaction:

\begin{equation}
    S_C = \frac{1}{2}\int_\q V_C(q) \rho_\q \rho_{-\q}\, ,
\end{equation}
where $V_C(q)$ is the gate-screened Coulomb potential defined in Eq.~\eqref{gatescr} and $\rho_\q$ is the density of electrons in the conduction bands as defined in Eq.~\eqref{density}. Diagrammatically, we will represent the Coulomb interaction between the electrons in the K-IVC conduction bands as  

\begin{equation}
\begin{tikzpicture}
\begin{feynman}
	\vertex (v1) at (0,1) {\(\alpha,\k+\q\)} ;
	\vertex (v2) at (0,-1) {\(\beta,\k\)};
	\vertex (v3) at (1,0) ;
	\vertex (v4) at (2,0) ;
	\vertex (v5) at (3,1) {\(\lambda,\k'-\q\)};
	\vertex (v6) at (3,-1) {\(\sigma,\k'\)};
	\vertex (v7) at (5.5,-0.1) {\(= - \left[F_\q(\bk)\right]_{\alpha\beta} V^C(q) \left[F_{-\q}(\k')\right]_{\lambda\sigma} \)};
	\diagram*{
		(v3) --[fermion] (v1),
		(v2) --[fermion] (v3),
		(v3) --[scalar] (v4),
		(v4) --[fermion] (v5),
		(v6) --[fermion] (v4)
	}; 
\end{feynman}
\end{tikzpicture}
\end{equation}

The third and fourth terms in Eq.~\eqref{actions} describe the coupling between the electrons and the Goldstone boson fields $\phi_\mu$. In particular, the third term is given by
\begin{equation}\label{int}
S_{\psi-\phi} = \int \mathrm{d}\tau\int_{\k,\q} g_{\alpha\beta}(\k,\q)\phi_\mu(\tau,\q) \bar{\psi}_{\alpha}(\tau,\k+\q)s^\mu\psi_{\beta}(\tau,\k)\, ,
\end{equation}
where the coupling $g_{\alpha\beta}(\k,\q)$ is defined in Eq.~\eqref{coupling}. As before, we represent the corresponding vertex diagrammatically as

\begin{equation}
\begin{tikzpicture}
\begin{feynman}
	\vertex (v1) at (-0.5,0) {\(q,\mu\)} ;
	\vertex (v2) at (1,0);
	\vertex (v3) at (2,1) {\(\alpha,s,\,k+q\)};
	\vertex (v4) at (2,-1) {\(\beta,s',\,k\)};
	\vertex (v5) at (3.5,-0.1) {\(=ig_{\alpha\beta}(\bk,\bq)[s^\mu]_{ss'}\, .\)};
	\diagram*{
		(v1) --[photon] (v2),
		(v2) --[fermion] (v3),
		(v4) --[fermion] (v2)
	}; 
\end{feynman}
\end{tikzpicture}
\end{equation}
The fourth term $S_{\psi-\phi^2}$ takes the form

\begin{equation}
    S_{\psi-\phi^2} = \frac 12 \,\int_{k,q,q'}\phi_\mu(q)\phi_\nu(q')  \bar{\psi}_{\alpha}(k+q+q')\left[\delta_{\mu\nu}\tilde{g}'_{\alpha\beta}(k,q,q') + (1-\delta_{\mu\nu})\bar{g}'_{\alpha\beta}(k,q,q')s^\mu s^\nu \right]\psi_\beta(k)\, ,
\end{equation}
where $k=(i\omega,\k)$, $q =(i\nu,\q)$, and $q'=(i\nu',\q')$ are three-vectors containing both frequency and momentum components, and $\int_k \equiv \int\frac{\mathrm{d}\omega}{2\pi}\int_\k$. The corresponding vertex is represented diagrammatically as

\begin{equation}
\begin{tikzpicture}
\begin{feynman}
	\vertex (v1) at (0,1) {\(q',\nu\)} ;
	\vertex (v2) at (0,-1) {\(q,\mu\)};
	\node [dot] (v3) at (1,0) ;
	\vertex (v4) at (2,1) {\(\alpha,s,\,k+q+q'\)};
	\vertex (v5) at (2,-1) {\(\beta,s',\, k\)};
	\vertex (v6) at (7.5,-0.1) {\(=-\frac{1}{2}\left[\delta_{\mu\nu}\tilde{g}'_{\alpha\beta}(\k,\q,\q')\delta_{ss'} + (1-\delta_{\mu\nu})\bar{g}'_{\alpha\beta}(\k,\q,\q')[s^\mu s^\nu]_{ss'}\right].\)};
	\diagram*{
		(v1) --[photon] (v3),
		(v2) --[photon] (v3),
		(v3) --[fermion] (v4),
		(v5) --[fermion] (v3)
	}; 
\end{feynman}
\end{tikzpicture}
\end{equation}
As before, only the coupling $\tilde{g}'_{\alpha\beta}$ will play a role in our discussion. The coupling function $\tilde{g}'_{\alpha\beta}(k,q,q')$ contains three different contributions. The first ``bare'' contribution comes from the second order term in the expansion of $e^{-i\hat{Q}}H_\Delta e^{i\hat{Q}}$, which, as discussed in Appendix~\ref{app:stiffness}, leads to the coupling $\tilde{g}_{\alpha\beta}(\k,\q,\q')$ defined in Eq.~\eqref{gtilde}. For future convenience, we point out that this coupling satisfies

\begin{eqnarray}
\tilde{g}_{\alpha\beta}(\k,0,0) & = & -\frac{1}{4}\langle u_{\alpha,\k}|\left[\tau^z,[\tau^z,\Delta(\k)]\right]\,|u_{\beta,\k}\rangle \label{gtildezero}\, .
\end{eqnarray}
The other two ``renormalization'' contributions to $\tilde{g}'_{\alpha\beta}(k,q,q')$ are the result of integrating out the valence electrons. The easiest way to represent these is to write out the different contributions to the coupling $\tilde{g}'_{\alpha\beta}(k,q,q')$ diagramatically as follows:

\begin{equation}\label{gbar}
\begin{tikzpicture}
\begin{feynman}
	\vertex (v1) at (0,0.75) ;
	\vertex (v2) at (0,-0.75) ;
	\node [dot] (v3) at (0.75,0) ;
	\vertex (v4) at (1.5,0.75) ;
	\vertex (v5) at (1.5,-0.75) ;
	\vertex (v6) at (1.8,0) {\(= \)} ;
	\diagram*{
		(v1) --[photon] (v3),
		(v2) --[photon] (v3),
		(v3) --[fermion] (v4),
		(v5) --[fermion] (v3)
	}; 
\end{feynman}
\end{tikzpicture}
\begin{tikzpicture}
\begin{feynman}
	\vertex (v1) at (0,0.75) ;
	\vertex (v2) at (0,-0.75) ;
	\vertex (v3) at (0.75,0) ;
	\vertex (v4) at (1.5,0.75) ;
	\vertex (v5) at (1.5,-0.75) ;
	\vertex (v6) at (1.8,0) {\(+\;\;\frac{1}{2}\)} ;
	\diagram*{
		(v1) --[photon] (v3),
		(v2) --[photon] (v3),
		(v3) --[fermion] (v4),
		(v5) --[fermion] (v3)
	}; 
\end{feynman}
\end{tikzpicture}
\begin{tikzpicture}
\begin{feynman}
	\vertex (v1) at (0,0.75) ;
	\vertex (v2) at (0,-0.75) ;
	\vertex (v3) at (0.75,0.6) ;
	\vertex (v33) at (0.75,-0.6) ;
	\vertex (v4) at (1.5,0.75) ;
	\vertex (v5) at (1.5,-0.75) ;
	\vertex (v6) at (1.8,0) {\(+\;\;\frac{1}{2} \)} ;
	\diagram*{
		(v1) --[photon] (v3),
		(v33) --[ghost, with arrow = 0.6 cm] (v3),
		(v2) --[photon] (v33),
		(v3) --[fermion] (v4),
		(v5) --[fermion] (v33)
	}; 
\end{feynman}
\end{tikzpicture}
\hspace{0.2 cm}
\begin{tikzpicture}
\begin{feynman}
	\vertex (v1) at (0,0.7) ;
	\vertex (v2) at (0,-0.7) ;
	\vertex (v3) at (0.75,0.6) ;
	\vertex (v33) at (0.75,-0.6) ;
	\vertex (v4) at (1.5,0.75) ;
	\vertex (v5) at (1.5,-0.75) ;
	\vertex (v6) at (1.5,0) {\(\, , \)} ;
	\diagram*{
		(v2) --[photon] (v3),
		(v33) --[ghost, with arrow = 0.6 cm] (v3),
		(v1) --[photon] (v33),
		(v3) --[fermion] (v4),
		(v5) --[fermion] (v33)
	}; 
\end{feynman}
\end{tikzpicture}
\end{equation}
where we have represented the propagator of the valence-band electrons by a dotted line with an arrow. The first diagram on the right hand side represents the ``bare'' coupling $\frac{1}{2}\tilde{g}_{\alpha\beta}(\k,\q,\q')$ discussed above. The last two diagrams on the right hand side correspond to the ``renormalization'' contributions involving a virtual valence band electron, with two vertices given by $g_{\alpha\beta}(\k,\q)$. Translating these diagrams into equations, the coupling $\tilde{g}'_{\alpha\beta}(k,q,q')$ is given by

\begin{align}
    \bar{g}_{\alpha\beta}(k,q,q') =  \tilde{g}_{\alpha\beta}(\k,\q,\q') +
    \sum_{\gamma<0} \left(\frac{g_{\alpha\gamma}(\k+\q,\q')g_{\gamma\beta}(\k,\q)}{i(\omega+\nu) - E_{\lambda,\k+\q}} + \frac{g_{\alpha\gamma}(\k+\q',\q)g_{\gamma\beta}(\k,\q')}{i(\omega+\nu') - E_{\lambda,\k+\q'}}  \right)\, ,
\end{align}
where the sum is over the K-IVC valence bands, labeled by negative integers.

The fourth term $S_\phi$ in Eq.~\eqref{actions} is the ``bare'' quadratic boson action obtained after integrating out the valence electrons and expanding the free energy up to the second order in $\phi$:

\begin{equation}\label{freeGoldstone}
S_{\phi} = \frac{1}{2}\int_q \phi_\mu(q) K_0(\q)  \phi_\mu(-q)\, ,
\end{equation}
where $K_0(\q)$ is given by the diagram

\begin{equation}\label{K0d}
\begin{tikzpicture}
\begin{feynman}
    \vertex (v0) at (-0.5,0.3) {\(K_0(\q) =  \)};
	\vertex (v1) at (0,-0.2) {\(\)};
	\vertex (v2) at (1,0.1);
	\vertex (v8) at (1,1);
	\vertex (v3) at (2,-0.2)  {\(\)};
	\diagram*{
	(v1) --[photon] (v2),
	(v2) --[ghost,half left,with arrow = 0.7 cm] (v8),
	(v8) --[ghost,half left,with arrow = 0.7 cm] (v2),
	(v2) --[photon] (v3);
	};
\end{feynman}
\end{tikzpicture}\, ,
\end{equation}
which evaluates to 

\begin{equation}\label{K0}
    K_0(\q) = 2\sum_{\alpha<0}\int\frac{\mathrm{d}\k}{(2\pi)^2} \tilde{g}_{\alpha\alpha}(\k,\q,-\q)\,.
\end{equation}
We would like to emphasize that the bare quantity $K_0(\q)$ at finite doping is very different from the quantity $K(\q)$ which we calculated in Appendix~\ref{app:stiffness} at charge neutrality. In particular, in general $K_0(0)$ will not be equal to zero, such that the boson $\phi$ appears to be massive. To obtain a proper, massless Goldstone mode propagator, we need to ``dress'' it with the RPA self-energy, which contains two terms. The first term originates from the $\phi\bar{\psi}\psi$ coupling and is given by

\begin{equation}\label{paraterm}
\begin{tikzpicture}
\begin{feynman}
    \vertex (v3) at (0.7,0)  {\(\Sigma_G^p(i\omega,\q) = \)};
	\vertex (v4) at (2,0);
	\vertex (v5) at (2.5,0);
	\vertex (v6) at (3.5,0);
	\vertex (v7) at (4,0);
	\diagram*{
	(v4) --[photon] (v5);
	(v5) --[fermion,half left](v6);
	(v6) --[fermion,half left] (v5);
	(v6) --[photon] (v7);
	};
\end{feynman}
\end{tikzpicture}\, .
\end{equation}
The second contribution to the boson self-energy comes from the $\phi^2\bar{\psi}\psi$ coupling and is given by

\begin{equation}\label{diaterm}
\begin{tikzpicture}
\begin{feynman}
    \vertex (v0) at (-1.2,0.3) {\(\Sigma_G^d(i\omega,\q) =  \)};
	\vertex (v1) at (0,-0.2) {\(\)};
	\node [dot] (v2) at (1,0.1);
	\vertex (v8) at (1,1);
	\vertex (v3) at (2,-0.2)  {\(\)};
	\diagram*{
	(v1) --[photon] (v2),
	(v2) --[fermion,half left] (v8),
	(v8) --[fermion,half left] (v2),
	(v2) --[photon] (v3);
	};
\end{feynman}
\end{tikzpicture}\, .
\end{equation}
Using the definition of the $\phi^2\bar{\psi}\psi$ coupling in Eq.~\eqref{gbar}, we can rewrite this as

\begin{align}
\begin{tikzpicture}
\begin{feynman}
	\vertex (v1) at (0.2,-0.2);
	\node [dot] (v2) at (1,0.15);
	\vertex (v8) at (1,1);
	\vertex (v3) at (1.8,-0.2);
	\vertex (v4) at (2,0.4) {\(=\)};
	\diagram*{
	(v1) --[photon] (v2),
	(v2) --[fermion,half left] (v8),
	(v8) --[fermion,half left] (v2),
	(v2) --[photon] (v3);
	};
\end{feynman}
\end{tikzpicture}
\begin{tikzpicture}
\begin{feynman}
	\vertex (v1) at (0.2,-0.2);
	\vertex (v2) at (1,0.15);
	\vertex (v8) at (1,1);
	\vertex (v3) at (1.8,-0.2);
	\vertex (v4) at (2,0.4) {\(+\)};
	\diagram*{
	(v1) --[photon] (v2),
	(v2) --[fermion,half left] (v8),
	(v8) --[fermion,half left] (v2),
	(v2) --[photon] (v3);
	};
\end{feynman}
\end{tikzpicture}
\begin{tikzpicture}
\begin{feynman}
	\vertex (v4) at (1,0);
	\vertex (v5) at (1.5,0);
	\vertex (v6) at (2.5,0);
	\vertex (v7) at (3.,0);
	\diagram*{
	(v4) --[photon] (v5);
	(v5) --[fermion,half left](v6);
	(v6) --[ghost,half left,with arrow=0.75 cm] (v5);
	(v6) --[photon] (v7);
	};
\end{feynman}
\begin{feynman}
    \vertex (v3) at (3.2,0)  {\(\;+ \)};
	\vertex (v4) at (3.5,0);
	\vertex (v5) at (4.,0);
	\vertex (v6) at (5.,0);
	\vertex (v7) at (5.5,0);
	\diagram*{
	(v4) --[photon] (v5);
	(v5) --[ghost,half left,with arrow=0.75 cm](v6);
	(v6) --[fermion,half left] (v5);
	(v6) --[photon] (v7);
	};
\end{feynman}
\end{tikzpicture}\, .
\end{align}
Evaluating all diagrams we find that the boson self-energy is given by

\begin{eqnarray}
\Sigma_G(i\omega,\q) & = & \Sigma^p_G(i\omega,\q) + \Sigma^d_G(i\omega,\q) \label{selfE} \\
 & = & 2\sum_{\alpha,\beta}\int\frac{\text{d}\k}{(2\pi)^2}\,\frac{n_{\alpha,\k+\q}-n_{\beta,\k}}{i\omega + E_{\alpha,\k+\q}-E_{\beta,\k}}|g_{\alpha\beta}(\k,\q)|^2  + 2\sum_{\alpha > 0} \int\frac{\text{d}\k}{(2\pi)^2}\,n_{\alpha,\k} \tilde{g}_{\alpha\alpha}(\k,\q,-\q)\, ,\nonumber
\end{eqnarray}
where the factors of two again come from spin degeneracy, and the indices $\alpha$ and $\beta$ in the first term run over both the K-IVC valence and conduction bands, i.e., $\alpha$ and $\beta$ run over both positive and negative integers. 

Including the self-energy, the ``dressed'' boson propagator equals

\begin{equation}
    D_R^{-1}(i\omega,\q) = -K_0(\q) - \Sigma_G(i\omega,\q) \, .
\end{equation}
We now claim that the properly ``dressed'' propagator does describe a massless boson, and therefore satisfies the following equation:

\begin{equation}\label{nomass}
    K_0(0) + \Sigma_G(0,0) = 0\, .
\end{equation}
To show that this condition is indeed satisfied, we start with using Eq.~\eqref{kzerog} to write

\begin{align}\label{massless1}
    \frac{n_{\alpha,\k}-n_{\beta,\k}}{E_{\alpha,\k}-E_{\beta,\k}}|g_{\alpha\beta}(\k,0)|^2 = 
    -\frac{1}{4}(n_{\alpha,\k}-n_{\beta,\k})\langle u_{\alpha,\k}|[\tau^z,\Delta(\k)]|u_{\beta,\k}\rangle \langle u_{\beta,\k} |\tau^z | u_{\alpha,\k}\rangle \, .
\end{align}
Note that because of the trace over the spin indices in the bubble diagrams, the screening corrections are only non-zero if the external boson line is labeled by $\mu = 0$. Summing over both $\alpha$ and $\beta$ in Eq.~\eqref{massless1}, we obtain the for the screened coupling to the $\mu = 0$ boson:

\begin{align}
 \sum_{\alpha,\beta}\frac{n_{\alpha,\k}-n_{\beta,\k}}{E_{\alpha,\k}-E_{\beta,\k}}|g_{\alpha\beta}(\k,0)|^2  = -\frac{1}{4}\big[\text{tr}(P_{\k}^o [\tau^z,\Delta(\k)]P_{\k}^u\tau^z) - \text{tr}(P_{\k}^u [\tau^z,\Delta(\k)]P_{\k}^o\tau^z) \big]\, ,
\end{align}
where $P_{\k}^o = \sum_{\alpha}n_{\alpha,\k} |u_{\alpha,\k}\rangle\langle u_{\alpha,\k}|$ is the projector onto the occupied states in the K-IVC bands at momentum $\k$, and $P_{\k}^u = \sum_{\alpha}(1-n_{\alpha,\k}) |u_{\alpha,\k}\rangle\langle u_{\alpha,\k}| $ is the projector onto the unoccupied states at $\k$. Using $P^u_\k = \mathds{1} - P^o_\k$, we find

\begin{equation}
 \sum_{\alpha,\beta}\frac{n_{\alpha,\k}-n_{\beta,\k}}{E_{\alpha,\k}-E_{\beta,\k}}|g_{\alpha\beta}(\k,0)|^2   = 
    \frac{1}{4}\text{tr}(P_{\k}^o [\tau^z,[\tau^z,\Delta(\k)]])  =   - \sum_{\alpha} n_{\alpha,\k} \tilde{g}_{\alpha\alpha}(\k,0,0)\, ,  \label{paramrew}
\end{equation}
where for the last equality we have used Eq.~\eqref{gtildezero}. Combining Eqs.~\eqref{paramrew}, \eqref{selfE}, and~\eqref{K0}, one finds that Eq.~\eqref{nomass} indeed holds, i.e., that $K_0(0)+\Sigma_G(0,0) = 0$. 

Since the bare boson propagator does not describe a massless Goldstone mode, it is crucial to always use the dressed propagator $D_R(i\omega,\q)$ to investigate the effect of the Goldstone modes on the conductive electrons. For small enough doping, the ``dressed'' propagator $D_R(i\omega,\q)$ will be very close to $D(i\omega,\q)$, the propagator of the Goldstone modes  at charge neutrality. This is because $D_R^{-1}(i\omega,\q) \approx \chi_R(\nu) (i\omega)^2 - \rho_{R}(\nu) \q^2$ changes continuously with doping and crosses over to $D^{-1}(i\omega,\q) \approx \chi_s (i\omega)^2 - \rho_s \q^2$ at $\nu=0$. We have numerically verified that $D^{-1}_R(i\omega,\q)$ at $\nu=1/4$ is indeed close to $D^{-1}(i\omega,\q)$. In the main text, we therefore use $D^{-1}_R(i\omega,\q) = D^{-1}(i\omega,\q)$ for simplicity. The main motivation for this is that we found the propagator obtained at charge neutrality to be less prone to numerical error. 

Using the propagator $D(i\omega,\q)$, the Goldstone mode-mediated interaction between the electrons is obtained from the following tree-level diagram:

\begin{equation}
\begin{tikzpicture}
\begin{feynman}
	\vertex (v1) at (0,1) {\( \alpha,\k+\q\)};
	\vertex (v2) at (0,-1) {\( \beta,\k\)};
	\vertex (v3) at (1,0) ;
	\vertex (v4) at (2,0) ;
	\vertex (v5) at (3,1) {\( \lambda,\k'-\q\)};
	\vertex (v6) at (3,-1) {\( \sigma,\k'\)};
	\vertex (v7) at (5,0) {\( = -V_{\alpha\beta\lambda\sigma}^G(i\omega,\q,\k,\k')\, ,\)};
	\diagram*{
		(v3) --[fermion] (v1),
		(v2) --[fermion] (v3),
		(v3) --[photon] (v4),
		(v4) --[fermion] (v5),
		(v6) --[fermion] (v4)
	}; 
\end{feynman}
\end{tikzpicture}
\end{equation}
where 

\begin{equation}
    V^G_{\alpha\beta\lambda\sigma}(i\omega,\q,\k,\k') = g_{\alpha\beta}(\k,\q)D(i\omega,\q)g_{\lambda\sigma}(\k',-\q)\, .
\end{equation} 
Equations~\eqref{GSint} and~\eqref{GSpotential} in the main text are then obtained by considering the static limit of $V^G$, i.e., setting $\omega=0$.

\section{Coulomb interaction and screening}\label{app:repulsive}

The bare repulsive interaction between the doped electrons in the K-IVC conduction bands is given by

\begin{equation}\label{coulomb}
H_C = \frac{1}{2A} \sum_\q V_C(q) :\rho_{\bq} \rho_{-\bq}:,
\end{equation}
where $V_C(q)$ the dual gate-screened Coulomb potential

\begin{equation}\label{gatescr}
V_C(q) = \frac{e^2}{2\epsilon_0\epsilon}\frac{\tanh(Dq)}{q}\, .
\end{equation}
In this expression, $D$ is the distance from the MATBG device to the metallic gates and $\epsilon$ is the dielectric constant. Note that in the Coulomb interaction, the sum over $\q$ is not restricted to the first mini-BZ, but goes over all BZ in the repeated zone scheme. The operators $\rho_\q$ are defined as

\begin{equation}\label{density}
\rho_{\bq} =  \sum_{\bk} \psi^\dagger_{\bk + \bq}F_{\bq}(\bk)  \psi_\bk
\end{equation}
and correspond to the density of electrons in the K-IVC conduction bands.  The form factors $F_\q(\k)$ which appear in this expression are given by

\begin{equation}
\left[F_\q(\k)\right]_{\alpha\beta} = \langle u_{\alpha,\k+\q}|\Lambda_{\q}(\k)|u_{\beta,\k}\rangle\, ,
\end{equation}
where $|u_{\alpha,\k}\rangle$ are the K-IVC eigenstates corresponding to the conduction bands and $\Lambda_\q(\k)$ are the form factors defined previously in Eq.~\eqref{formfactor}. The form factors $\Lambda_{\bq}(\bk)$ result from expressing the Coulomb interaction in the BM band basis (see, e.g., Ref.~\cite{KIVC} for details). Note that because of these form factors the Coulomb interaction acquires an explicit dependence on the incoming momenta $\k$ and $\k'$. 

Because there is a Fermi surface at $\nu = 1/4$, the electrons can efficiently screen the Coulomb interaction. To take this effect into account, we calculate the standard (static) polarization bubble, which evaluates to

\begin{equation}\label{Pi}
\Pi(\q) = -2\int \frac{\text{d}\k}{(2\pi)^2}\sum_{\alpha,\beta}\frac{n_{\alpha,\k+\q}-n_{\beta,\k}}{E_{\alpha,\k+\q} - E_{\beta,\k}}\;\big|\left[F_\q(\k)\right]_{\alpha\beta}\big|^2\, ,
\end{equation}
where $n_{\alpha,\k}=n(E_{\alpha,\k}) = \Theta(\varepsilon_F - E_{\alpha,\bk})$ is the zero-temperature Fermi-Dirac distribution ($\Theta(x)$ is the Heaviside step function) representing the fermion occupation numbers, $\varepsilon_F$ is the Fermi energy, and the factor of two comes from the spin degeneracy. 

From the polarization bubble we obtain the dielectric function $\epsilon(\q) = 1 + V_C(q)\Pi(\q)$, which appears in the static RPA screened Coulomb interaction

\begin{eqnarray}\label{scrCoulomb}
V_{\alpha\beta\lambda\sigma}^{C,scr}(\q,\k,\k') & = & \frac{V_C(q)}{\epsilon(\q)} \left[F_\q(\k)\right]_{\alpha\beta} \left[F_{-\q}(\k')\right]_{\lambda\sigma} \\
 & = & \frac{e^2}{2\epsilon_0\epsilon}\frac{\tanh(Dq)}{q + k_s(\q)}\left[F_\q(\k)\right]_{\alpha\beta} \left[F_{-\q}(\k')\right]_{\lambda\sigma}\, . \nonumber
\end{eqnarray}
In the last line, we have defined

\begin{equation}
k_s(\q) = \frac{e^2}{2\epsilon_0\epsilon} \Pi(\q)\tanh(Dq)\, , \label{Eq:ksdef}
\end{equation}
which is a $\q$-dependent inverse screening length.

\section{Interactions in the Cooper channel and Kramers time-reversal symmetry}\label{app:Kramers}

The goal of this appendix is to show that as a result of the Kramers time-reversal symmetry of the doped K-IVC state, the Goldstone-mediated and Coulomb interactions in the Cooper channel can be written as in Eqs. \eqref{positive1} and \eqref{positive2} in the main text. For the Coulomb interaction, this means that we will show that it can be written as follows:

\begin{eqnarray}
\hat{V}^C_{\alpha\beta}(\k',\k) & \equiv & \sum_\G \frac{e^2}{2\epsilon_0\epsilon}\frac{\tanh(D|\k'-\k+\G|)}{|\k'-\k+\G| + k_s(\k'-\k+\G)} \left[F_{\k'-\k+\G}(\k)\right]_{\alpha\beta} \left[F_{\k-\k'-\G}(-\k)\right]_{\alpha\beta} \nonumber \\
 & = & e^{i\varphi_\alpha(\k')}\left(\sum_\G \frac{e^2}{2\epsilon_0\epsilon}\frac{\tanh(D|\k'-\k+\G|)}{|\k'-\k+\G| + k_s(\k'-\k+\G)} \big|\left[F_{\k'-\k+\G}(\k)\right]_{\alpha\beta}\big|^2 \right)e^{-i\varphi_\beta(\k)}\, , \label{tr}
\end{eqnarray}
where the sum is over moir\'e reciprocal lattice vectors $\G$ and $e^{i\varphi_{\alpha}(\bk')}$, $e^{-i\varphi_{\beta}(\bk)}$ are gauge-dependent phase factors.

We start by recalling the definition of the K-IVC form factors $F_\q(\k)$:

\begin{equation}
\left[F_{\q}(\k)\right]_{\alpha\beta}= \langle u_{\alpha,\k+\q}|\Lambda_\q(\k)|u_{\beta,\k}\rangle\, ,
\end{equation}
where $|u_{\alpha,\k}\rangle$ are the K-IVC eigenstates. Because of the spinless time-reversal symmetry $\mathcal{T}=\tau_x K$ of the BM model, we can without loss of generality use BM form factors which satisfy $\Lambda_{-\q}(-\k) = \tau_x \Lambda^*_\q(\k)\tau_x$ and $\tau^z \Lambda_\q(\k)\tau^z = \Lambda_\q(\k)$. Using these properties, we find that

\begin{equation}
\left[F_{-\q}(-\k)\right]_{\alpha\beta} = \langle u_{\alpha,-\k-\q}|i\tau_y \Lambda^*_\q(\k)i\tau_y^T|u_{\beta,-\k}\rangle
\end{equation}
Because of the $\mathcal{T}'$ symmetry, with $\mathcal{T}' = \tau^z \mathcal{T} = i \tau_y K$, the K-IVC eigenstates satisfy

\begin{equation}\label{Tbloch}
\mathcal{T}'|u_{\beta,\k}\rangle = e^{i\varphi_\beta(\k)} |u_{\beta,-\k}\rangle \Rightarrow i\tau_y^T|u_{\beta,-\k}\rangle = e^{-i\varphi_\beta(\k)}|u_{\beta,\k}\rangle^*\, ,
\end{equation}
where $e^{i\varphi_\beta(\k)}$ is a gauge-dependent phase factor. Since $|u_{\alpha,\k+\G}\rangle = |u_{\alpha,\k}\rangle$, it follows that $e^{i\varphi_\alpha(\k+\G)} = e^{i\varphi_\alpha(\k)}$. Using Eq.~\eqref{Tbloch}, one finds that the K-IVC form factors satisfy

\begin{equation}
\left[F_{-\q}(-\k)\right]_{\alpha\beta} = e^{i[ \varphi_\alpha(\k+\q) -\varphi_\beta(\k)]} \langle u_{\alpha,\k+\q}| \Lambda_\q(\k)|u_{\beta,\k}\rangle^*  = e^{i[\varphi_\beta(\k) - \varphi_\alpha(\k+\q)]} \left[F_{\q}(\k)\right]^*_{\alpha\beta} \, ,
\end{equation}
which in turn implies Eq.~\eqref{tr}.

Next, we show that the Goldstone-mediated interaction in the Cooper channel similarly satisfies

\begin{eqnarray}
\hat{V}^G_{\alpha\beta}(\k,\k') & \equiv & -\sum_\G K(\k'-\k+\G)^{-1}g_{\alpha\beta}(\k,\k'-\k+\G) g_{\alpha\beta}(-\k,\k-\k'-\G) \nonumber \\
 & = & -e^{i\varphi_\alpha(\k')}\left(\sum_\G K(\k'-\k+\G)^{-1}\big|g_{\alpha\beta}(\k,\k'-\k+\G)\big|^2\right)e^{-i\varphi_\beta(\k)}. \label{IntGT}
\end{eqnarray}
First, we again use the properties of the BM form factors and find 

\begin{eqnarray}
g_{\alpha\beta}(-\k,-\q) & = & \frac{i}{2}\langle u_{\alpha,-\k-\q}|\Delta(-\k-\q)i\tau_y\Lambda^*_{\q}(\k)i\tau_y^T \tau^z-\tau^z i\tau_y\Lambda^*_\q(\k)i\tau_y^T\Delta(-\k)|u_{\beta,-\k}\rangle.
\end{eqnarray}
The Kramers time-reversal symmetry of the K-IVC state implies that $i\tau_y^T \Delta(-\k)i\tau_y = \Delta^*(\k)$, which allows us to write

\begin{eqnarray}
g_{\alpha\beta}(-\k,-\q) & = & -\frac{i}{2}\langle u_{\alpha,-\k-\q}|i\tau_y\big[\Delta^*(\k+\q)\Lambda^*_{\q}(\k)\tau^z -\tau^z \Lambda^*_\q(\k)\Delta^*(\k)\big] i\tau_y^T|u_{\beta,-\k}\rangle.
\end{eqnarray}
From the transformation property of the K-IVC states in Eq.~\eqref{Tbloch}, we find that

\begin{equation}
    g_{\alpha\beta}(-\k,-\q) = e^{i[\varphi_\alpha(\k+\q) - \varphi_\beta(\k)]} g_{\alpha\beta}^*(\k,\q)\, ,
\end{equation}
which implies Eq.~\eqref{IntGT}.

\section{BCS gap equation}\label{app:BCS}

In the main text, we look for superconducting states with an order parameter of the form

\begin{eqnarray}
\tilde{{\boldsymbol{\Delta}}}_\k^t & = & \left(\begin{matrix} \tilde{  \Delta}_{1,\k} & 0 \\ 0 & \tilde{  \Delta}_{2,\k} \end{matrix}\right)\otimes is_y{\boldsymbol s}\, ,
\end{eqnarray}
corresponding to spin-triplet pairing of electrons within the same band. The finite temperature gap equation is then given by

\begin{equation}\label{gapeq}
\tilde{\Delta}_{\alpha,\k} = -\frac{1}{A}\sum_{\beta,\k'} V_{\alpha\beta}(\k,\k') \frac{\tilde{\Delta}_{\beta,\k'}}{2\sqrt{E_{\beta,\k'}^2 + |\tilde{\Delta}_{\beta,\k'}|^2}}\tanh\left(\frac{\sqrt{E_{\beta,\k'}^2 + |\tilde{\Delta}_{\beta,\k'}|^2}}{2k_BT}\right)\,,
\end{equation}
where $A$ is the area of the system. To find the critical temperature of possible superconducting states, we take the BCS gap equation and apply the standard procedure by ignoring the dependence on the gap in the denominator and the argument of the hyperbolic tangent in Eq.~\eqref{gapeq}, motivated by the fact that the gap goes to zero if we approach the critical temperature from below. This leaves us with

\begin{eqnarray}\label{gapeq2}
\tilde{\Delta}_{\alpha,\k} & = & -\frac{1}{A}\sum_{\beta,\k'} V_{\alpha\beta}(\k,\k') \frac{ \tilde{\Delta}_{\beta,\k'} }{ 2E_{\beta,\k'} }\tanh\left(\frac{E_{\beta,\k'}}{2k_BT_c}\right) \nonumber \\
 & \approx & - \sum_\beta \int \mathrm{d}E\; N_\beta(E)\int\frac{\mathrm{d}\theta'}{2\pi} V_{\alpha\beta}[\k,\k'(E,\theta')] \frac{ \Delta_{\beta,\k'(E,\theta')} }{ 2E }\tanh\left(\frac{E}{2k_BT_c}\right)\, ,\label{isotropic}
\end{eqnarray}
where $\theta'$ is a polar angle in momentum space and

\begin{equation}
    N_\beta(E) = \int \frac{\mathrm{d}\theta}{2\pi} \frac{k_{\beta}(E,\theta)}{2\pi}\left| \frac{\partial E_{\beta}(k,\theta)}{\partial k} \right|^{-1}_{k = k_{\beta}(E,\theta)}
\end{equation}
corresponds to the density of states in band $\beta$. Here, we use $k_\beta(E,\theta)$ to denote the inverse function of the dispersion $E_\beta(k,\theta)$, i.e., it is defined via the relation $E_\beta[k_\beta(E',\theta),\theta] = E'$. Note that the approximation in Eq.~\eqref{isotropic} is justified if the  dispersion near the Fermi surfaces is close to being isotropic.

Next, focusing on the vicinity of the Fermi surface, we write

\begin{eqnarray}
\tilde{\Delta}_{\alpha,\k} & \approx & - \sum_\beta N_\beta(0) \int\frac{\mathrm{d}\theta'}{2\pi} V_{\alpha\beta}[\k,\k_{F,\beta}(\theta')] \tilde{\Delta}_{\beta,\k_{F,\beta}(\theta')}  \int  \mathrm{d}E\; \frac{ \tanh\left(\frac{E}{2k_BT_c}\right) }{ 2E } \nonumber \\
\Rightarrow \tilde{\Delta}_\alpha(\theta) & \approx  &- \sum_\beta N_\beta(0) \int\frac{\mathrm{d}\theta'}{2\pi} V_{\alpha\beta}(\theta,\theta') \tilde{\Delta}_{\beta}(\theta')  \int  \mathrm{d}E\; \frac{ \tanh\left(\frac{E}{2k_BT_c}\right) }{ 2E }\, ,
\end{eqnarray}
where $\k_{F,\beta}(\theta')$ is the angle-dependent Fermi momentum on the Fermi surface in band $\beta$. In the last line, we have introduce the notation $\Delta_{\beta}(\theta) = \Delta_{\beta,\k_{F,\alpha}(\theta)}$ for the gaps on the Fermi surfaces, and also $V_{\alpha\beta}(\theta,\theta') = V_{\alpha\beta}[\k_{F,\alpha}(\theta),\k_{F,\beta}(\theta')]$ for the interaction on the Fermi surfaces. $N_\beta(0)$ is the density of states of band $\beta$ at the Fermi surface.

It is now clear that to find solutions of the gap equation, we have to solve the eigenvalue equation

\begin{equation}
 \sum_\beta \int\frac{\mathrm{d}\theta'}{2\pi} V_{\alpha\beta}(\theta,\theta')N_\beta(0) \tilde{\Delta}_{\beta}(\theta') = -\lambda \tilde{\Delta}_{\alpha}(\theta)\, ,
\end{equation}
after which we can proceed with the solution of the gap equation in the standard way to obtain

\begin{equation}
    k_BT_c \sim \varepsilon_F\times e^{-1/\lambda}\, ,
\end{equation}
where $\varepsilon_F$ is the Fermi energy.

\section{Continuous transition between K-IVC and valley Hall insulator}\label{app:VH}

If the hexagonal boron-nitride (hBN) substrate encapsulating the MATBG system becomes sufficiently aligned with one of the graphene layers, it can introduce a significant $C_{2z}$-breaking sublattice splitting $\Delta_t \sigma_z$ via the proximity effect \cite{Jung,Kim2018,Zibrov,YankowitzJung}. Here, $\sigma_i$ are the Pauli matrices acting on the sublattice index. The sublattice splitting generates a Dirac mass at both the $K$ and $K'$ points of the mini-BZ, leading to an insulating single-particle spectrum at charge neutrality. Depending on the sign of the sublattice splitting, the spin resolved bands in each valley have Chern number $\pm 1$ \cite{Zou2018,AHpaper,ZhangMao}. Note that time-reversal symmetry is not broken and that the bands in different valleys which are exchanged under time-reversal have opposite Chern numbers. Because only the valley-resolved Chern number of the filled bands is non-zero, this state is referred to as the valley Hall (VH) insulator. 

In Ref.~\cite{KIVC} it was found that within mean-field theory, there is a transition from the K-IVC insulator to the VH insulator at a critical sublattice splitting (say, on the top layer) of $\Delta_t^* \sim 10$ meV. At this point, there is a second order phase transition where both the $U_V(1)$ and time-reversal symmetry are restored. Importantly, the single-particle gap does not close at the transition. We also find that the Fermi surfaces around $\Gamma$ at $\nu = 1/4$ do not change in any significant way if we tune through the transition. However, if $\Delta_t$ becomes sufficiently close to the critical value $\Delta_t^*$, there is an additional soft (critical) bosonic mode which can facilitate pairing. 

Based on experience with other systems with a Fermi surface coupled to a critical mode one naturally expects non-Fermi liquid behavior near the K-IVC -- VH transition, even at small coupling \cite{RMPRoschVojta,RMPStewart,Oganesyan,Andergassen,AbanovChubukov,Metlitski1,Metlitski2,SSLee}. We will argue that this expectation is essentially correct, but also that the non-Fermi liquid physics follows from very small, seemingly negligible terms. To set up the argument, let us actually start from the VH side, i.e., let us consider the system with $\Delta_t>\Delta^*_t$. Also, in this Appendix, we will work in the sublattice polarized basis introduced in Ref.~\cite{KIVC}. As the precise definition of this basis is not relevant for this work, we will not give it here and just refer to Ref.~\cite{KIVC} for details. The only reason why we use the sublattice polarized basis is that the K-IVC order parameter takes on a particularly simple form. Namely, in this basis we have $\Delta(\k) = [d_x(\k)\tau_x+d_y(\k)\tau_y]\sigma_y$, where $\tau_i$ are still the Pauli matrices acting on the valley index. The Yukawa coupling to the K-IVC modes which become soft near the critical point is then given by 
 
\begin{equation}\label{yuk}
H_{Y} = \frac{g}{N}\sum_{\k,\q}\tilde{c}^\dagger_{\k+\q}(\vv{\phi}_\q \cdot \vv{\tau})\sigma_y \tilde{c}_\k\, ,
\end{equation}
where $\vv{\phi}_\q \cdot \vv{\tau}=\phi^x_\q\tau_x + \phi^y_\q\tau_y$. So the doped VH state coupled to the K-IVC modes is described by

\begin{equation}
H = \sum_{\k,\tau,\alpha} \varepsilon_{\tau,\alpha,\k}\tilde{c}^\dagger_{\tau,\alpha,\k}\tilde{c}_{\tau,\alpha,\k} +
 \frac{1}{N}\sum_{\k,\q}\sum_{\tau,\tau',\alpha,\beta} \langle w_{\tau,\alpha,\k+\q}|(\vv{\phi}_\q \cdot \vv{\tau})\sigma_y|w_{\tau',\beta,\k}\rangle \tilde{c}^\dagger_{\tau,\alpha,\k+\q}\tilde{c}_{\tau',\beta,\k},
\end{equation}
where $\varepsilon_{\tau,\alpha,\k}$ and $|w_{\tau,\alpha,\k}\rangle$ are the band energies and Bloch states of the mean-field VH Hamiltonian. Because the VH state preserves the valley symmetry, the eigenstates have a well-defined valley quantum number $\tau$. The index $\alpha$ distinguishes between valence and conduction bands. Since we are interested in, e.g., electron doping, we neglect the valence bands and focus only on the conduction bands. This means that we can ignore the $\alpha$ index, and label the electrons by the valley index (and spin). We can then write the Hamiltonian as

\begin{equation}
H = \sum_{\k,\tau} \varepsilon_{\tau,\k}\tilde{c}^\dagger_{\tau,\k}\tilde{c}_{\tau,\k}
 + \frac{1}{N}\sum_{\k,\q}\left(g_{+-}(\k,\q)\phi^+_\q \tilde{c}^\dagger_{+,\k+\q}\tilde{c}_{-,\k} + H.c. \right)\, ,
\end{equation}
where in the last line we have introduced the notation $\phi^+_\q = \phi^x_\q+i\phi^y_\q$  and

\begin{equation}\label{gpm}
g_{+-}(\k,\q) = g\langle w_{+,\k+\q}|\tau_x\sigma_y|w_{-,\k}\rangle.
\end{equation}
For our purposes, the main question we want to address is whether the coupling $g_{+-}(\k,\q)$ becomes zero at zero momentum transfer, i.e., whether $g_{+-}(\k,0)=0$ or not.

Before we answer the above question, we first recall that the BM model has an emergent approximate particle-hole symmetry $\mathcal{P}$~\cite{SongWang,Hejazi}, which acts in the sublattice polarized basis as~\cite{KIVC}

\begin{equation}
    \mathcal{P} \;\;:\;\;\tilde{c}^\dagger_{\k} \rightarrow \tau^z\sigma_y \tilde{c}_{-\k}.
\end{equation}
If we combine the particle-hole symmetry with the time-reversal symmetry $\mathcal{T}$ defined in Eq.~\eqref{Tsymm}, we obtain an approximate $\mathcal{PT}$ symmetry acting as

\begin{equation}
    \mathcal{PT}\;\;:\;\;\tilde{c}^\dagger_{\k} \rightarrow i\tau_y\sigma_y\tilde{c}_{\k}\;\;,\;i\rightarrow -i.
\end{equation}
Because of this approximate $\mathcal{PT}$ symmetry we conclude that the dominant, particle-hole symmetric terms in the VH Hamiltonian anti-commute with $\tau_y\sigma_y$, and therefore also with $\tau_x\sigma_y$. But these matrices exactly constitute the K-IVC order parameter in the sublattice polarized basis. This implies that if the VH Hamiltonian was perfectly particle-hole symmetric, then the coupling would vanish for zero momentum transfer: $g_{+-}(\k,0) = 0$. This is because $\tau_x\sigma_y$ anti-commutes with the particle-hole symmetric VH Hamiltonian, such that it maps a conduction band Bloch state to a valence band Bloch state and vice versa. Because of this, the matrix element in Eq.~\eqref{gpm} is strictly zero. Note that the full Yukawa coupling term in Eq.~\eqref{yuk} is not zero when $\q=0$, only the part projected onto the conduction bands is. In other words, at $\q=0$, the Yukawa coupling only mixes the valence and conduction bands of the particle-hole symmetric VH insulator, but it does not mix conduction bands among themselves. In part, this is a manifestation of the fact that the hBN sublattice splitting $\sigma_z$ and the K-IVC order parameter $\tau_x\sigma_y$ anti-commute, which also implies that the electron gap at $\nu=0$ does not close at the critical point.

In general, we of course have to include the small particle-hole symmetry breaking terms in the VH Hamiltonian. These terms give rise to a non-zero, but very small value for $g_{+-}(\k,0)$. 

Similar to the analysis in Ref.~\cite{Lederer}, we can now consider the electron interaction induced by the soft K-IVC modes for $\Delta_t \gtrsim \Delta_t^*$. It is given by~\cite{Lederer}

\begin{equation}\label{KIVCint}
V_{IVC}(\k,\q,\omega) = -|g_{+-}(\k,\q)|^2 \chi(\q,\omega)\,.
\end{equation}
Here, $\chi(\q,\omega)$ is the valley-$U(1)$ susceptibility

\begin{equation}
\chi(\q,\omega)\sim \chi_0\left(\frac{\xi^{-2} }{c^2q^2+\omega^2+ \xi^{-2}}\right)^{1-\eta/2}\, ,
\end{equation}
where $\xi\sim |\Delta_t-\Delta_t^*|^{-\nu} = |\delta \Delta_t|^{-\nu}$ is the correlation length of the boson field $\phi$ and $\chi_0 \sim |\delta \Delta_t|^{-\gamma}$. The critical exponents $\nu,\gamma$, and $\eta$ are those of the (2+1)-$d$ $O(2)$ Wilson-Fisher fixed point \cite{WilsonFisher}.

Because $g_{+-}$ is non-zero at $\q=0$, the interaction $V_{IVC}(\k,0,0)$ in Eq.~\eqref{KIVCint} diverges at the critical point, resulting in non-Fermi liquid physics. However, because $g_{+-}(\k,0)$ is very small, we only expect the non-Fermi liquid physics to manifest itself at very long distance and time scales.

\end{appendix}

\end{document}